\documentclass[preprint,12pt,3p]{elsarticle}

\usepackage{amssymb}
\newdefinition{rmk}{Remark}
\usepackage{hyperref}
\usepackage{graphicx}

\usepackage{epstopdf} % make WinEdt able to compile eps file by using the default compile button!

\usepackage{caption}
\DeclareGraphicsExtensions{.eps,.mps,.pdf,.jpg,.png}
\graphicspath{{figures/}{../figures/}}

\usepackage{fancyhdr}

\usepackage{amsmath}
\usepackage{amsfonts}

\usepackage{dsfont}

\usepackage{array}
\newcolumntype{C}[1]{>{\centering\let\newline\\\arraybackslash\hspace{0pt}}m{#1}}

\newcommand{\dd}{{\rm d}}
\newcommand{\kB}{k_{\mathrm{B}}}
\newcommand{\Ar}{\mathrm{A}}
\newcommand{\Br}{\mathrm{B}}

\newcommand{\Or}{\mathrm{O}}
\newcommand{\Tr}{\mathrm{Tr}}
\newcommand{\E}{\mathbb{E}}
\newcommand{\Var}{\mathrm{Var}}

\newcommand{\Lg}{\mathcal{L}}
\newcommand{\PP}{\overline{\mathbf{P}}}
\newcommand{\TT}{\mathrm{T}}

\newcommand{\e}{{\mathbf{e}}}
\newcommand{\x}{{\mathbf{x}}}

\newcommand{\q}{{\mathbf{q}}}
\newcommand{\p}{{\mathbf{p}}}

\newcommand{\A}{{\mathbf{A}}}
\newcommand{\B}{{\mathbf{B}}}

\newcommand{\F}{{\mathbf{F}}}
\newcommand{\G}{{\mathbf{G}}}
\newcommand{\M}{{\mathbf{M}}}
\newcommand{\I}{{\mathbf{I}}}

\newcommand{\D}{{\mathrm{D}}}
\newcommand{\R}{{\mathrm{R}}}

\newcommand{\WB}{{\mathrm{\mathbf{W}}}}

\newcommand{\GammaB}{{\boldsymbol{\Gamma}}}

\newcommand{\SigmaB}{{\boldsymbol{\Sigma}}}

\newtheorem{proposition}{Proposition}

\usepackage{mathrsfs}

\usepackage{soul}
\usepackage{color}
\usepackage[titletoc,toc,title]{appendix}

\usepackage{mathdots}

\begin{document}

\begin{frontmatter}

\title{Stochastic Norton dynamics: An alternative approach for the computation of transport coefficients in dissipative particle dynamics}

\author[label1]{Xinyi Wu}
\address[label1]{School of Mathematics, University of Birmingham, Edgbaston, Birmingham, B15 2TT, United Kingdom}
\ead{xxw142@student.bham.ac.uk}
%\ead[url]{https://sites.google.com/site/duonghongmath/}

\author[label1]{Xiaocheng Shang\corref{cor1}}

\ead{x.shang.1@bham.ac.uk}
%\ead[url]{https://sites.google.com/site/xiaochengshang/, https://www.birmingham.ac.uk/staff/profiles/maths/shang-xiaocheng.aspx}

\cortext[cor1]{Corresponding author.}

\begin{abstract}
  We study a novel alternative approach for the computation of transport coefficients at mesoscales. While standard nonequilibrium molecular dynamics (NEMD) approaches fix the forcing and measure the average induced flux in the system driven out of equilibrium, the so-called ``stochastic Norton dynamics'' instead fixes the value of the flux and measures the average magnitude of the forcing needed to induce it. We extend recent results obtained in Langevin dynamics to consider the generalisation of the stochastic Norton dynamics in the popular dissipative particle dynamics (DPD) at mesoscales, important for a wide range of complex fluids and soft matter applications. We demonstrate that the responses profiles for both the NEMD and stochastic Norton dynamics approaches coincide in both linear and nonlinear regimes, indicating that the stochastic Norton dynamics can indeed act as an alternative approach for the computation of transport coefficients, including the mobility and the shear viscosity, as the NEMD dynamics. In addition, based on the linear response of the DPD system with small perturbations, we derive a closed-form expression for the shear viscosity, and numerically validate its effectiveness with various types of external forces. Moreover, our numerical experiments demonstrate that the stochastic Norton dynamics approach clearly outperforms the NEMD dynamics in controlling the asymptotic variance, a key metric to measure the associated computational costs, particularly in the high friction limit.
\end{abstract}


\end{frontmatter}

% \linenumbers

\pagenumbering{arabic}
% =========================================================
% =========================================================

\section{Introduction}
\label{sec:Introduction}

Molecular dynamics is a computer simulation method for analysing the physical movements of atoms and molecules, and has been widely used in chemical physics, materials science, and biophysics~\cite{Allen2017,Frenkel2001,Leimkuhler2015b}. In classical molecular dynamics, the equations of motion of individual atoms and molecules are governed by Newton's law in the microcanonical ensemble, where the energy (i.e., the Hamiltonian of the system) is conserved. However, the constant energy setting is not ideal for a real-world laboratory environment since energy, as an extensive variable, is dependent on the size of the system. Instead, the canonical ensemble, where the temperature is conserved using suitable ``thermostat'' techniques, is typically used. 

One popular thermostat is the Langevin dynamics, commonly used for molecular dynamics simulations. In Langevin dynamics, in addition to the conservative forces, each particle is subject to dissipative and collisional interactions with the particles of an artificial ``heat bath''. Langevin dynamics supplements the conservative Newtonian equations of motion with balanced damping and stochastic terms in such a way that the desired target system temperature is maintained. Langevin dynamics has been widely applied in a range of applications in molecular dynamics~\cite{Allen2017,Frenkel2001,Leimkuhler2015b}; more recently, variants of Langevin dynamics (e.g., adaptive Langevin dynamics) have attracted increasing attention in data science~\cite{Leimkuhler2015a,Shang2015,Lei2016a,Leimkuhler2019a,Duong2021,Sekkat2023}.

However, as previously discussed in physics literature~\cite{Leimkuhler2016a}, Langevin dynamics is not suitable for simulations where hydrodynamic interactions are of interest. In order to be consistent with hydrodynamics, a particle model should respect Galilean invariance, and particularly should conserve momentum~\cite{Hoogerbrugge1992}; however, the momentum is not conserved in Langevin dynamics. Moreover, due to the violation of global momentum conservation, Langevin dynamics can lead to nonphysical screening of hydrodynamic interactions~\cite{Duenweg1993}. Therefore, when hydrodynamic interactions are important, standard thermostats should be replaced by momentum-conserving thermostats, in particular the so-called dissipative particle dynamics (DPD) method proposed by Hoogerbrugge and Koelman~\cite{Hoogerbrugge1992}.

In DPD, each particle is regarded as a coarse-grained particle consisting of several molecules. In its traditional formulation, these DPD particles interact in a soft (and short-ranged) potential, which allows larger integration timesteps than would be possible in molecular dynamics. Simultaneously, the coarse graining nature of DPD particles means that the number of degrees of freedom required is (significantly) decreased. It is worth mentioning that alternative potentials have also been widely used in DPD, including the widely used Lennard-Jones potential in polymer melts~\cite{Shang2017}. As in Langevin dynamics, a thermostat consisting of well-balanced damping and stochastic terms is applied to each particle in DPD. However, unlike Langevin dynamics, both damping and stochastic terms in DPD are pairwise. Given also that the damping term is based on relative velocities, the formulation of DPD leads to the conservation of both the angular momentum and the linear momentum, essential for the preservation of the hydrodynamics. Moreover, the dependence on the relative velocity makes DPD a profile-unbiased thermostat (PUT)~\cite{Evans1986,Evans2008} by construction and therefore it is an ideal thermostat for nonequilibrium molecular dynamics~\cite{Soddemann2003}. Due to the aforementioned properties, DPD has been widely used in a large number of complex fluids and soft matter applications, including colloids~\cite{Koelman1993}, blood~\cite{Fedosov2013}, and polymers~\cite{Spenley2000} (see more applications in an excellent review~\cite{Espanol2017} and references therein).

% Moreover, the theoretical analysis for shear viscosity computation in Brownian motions can be found in~\cite{Pavliotis2008a}, while Pavliotis and Vogiannou focus on studying underdamped and overdamped dynamics.

One particular challenge in Langevin dynamics and DPD, and molecular dynamics in general, is the computation of transport coefficients (e.g., mobility, diffusion coefficient, and shear viscosity) in an accurate and efficient manner. Transport coefficients measure how rapidly a perturbed system returns to equilibrium. One standard approach in equilibrium is the celebrated Green--Kubo formulas~\cite{Green1954,Kubo1957} based on integrals of suitable time correlation functions~\cite{Leimkuhler2015,Shang2019}. For example, the shear viscosity can be calculated by integrating the stress-stress autocorrelation function (see examples in the DPD literature~\cite{Chaudhri2010,Panoukidou2021}). However, time correlation functions represent the average response to the naturally occurring (therefore very small) fluctuations in the system properties. Given that the equilibrium stress is subject to large fluctuations in measuring the shear viscosity~\cite{Backer2005}, the signal-to-noise ratio is particularly unfavorable at long times, where there may be a significant (but unrelated) contribution to the integral defining a transport coefficient~\cite{Allen2017,Palmer1994}. Therefore, the simulation must be run for long times in order to have highly converged values for the Green--Kubo integrand over the range of times which contribute significantly to the integral~\cite{Palmer1994}.

Another popular approach is the so-called nonequilibrium molecular dynamics (NEMD) method, where the steady state of the system is subject to external perturbations (either stationary fluxes or spatial gradients of some quantities), in order to artificially induce larger fluctuations, thereby improving the signal-to-noise ratio of the measured response~\cite{Evans2008,Joubaud2012}. By measuring the steady state response to such a perturbation, the decay to the equilibrated state is then related to the corresponding transport coefficient, which avoids problems with long-time behaviour of correlation functions~\cite{Allen2017}. However, estimators derived from these approaches also suffer from large statistical errors, as quantified in~\cite[Proposition 2.4]{Spacek2023}. As a consequence, convergence requires the simulation of very long trajectories, resulting in high computational cost. To this end, a key metric to measure this cost is, for any given method, the asymptotic variance of estimators of the transport coefficient. Although variance
reduction techniques have been proposed to reduce the variance in the computation of transport coefficients (see~\cite{Stoltz2022} for a recent review), efficiently estimating these quantities is a challenging but still important area of research.

While standard NEMD approaches fix the forcing and measure the average induced flux in the system driven out of equilibrium, the so-called ``Norton dynamics'', studied by Evans and Morriss in the 1980s, instead fixes the value of the flux and measures the average magnitude of the forcing needed to induce it~\cite{Evans1985}. While showing promising results, the Norton dynamics was only considered in a deterministic setting. Recently, Blassel and Stoltz extended the Norton dynamics to a stochastic setting~\cite{Blassel2023}. They first developed a general formal theory for a broad class of diffusion processes, before specialising it to Langevin dynamics. Moreover, they provided numerical evidence that the ``stochastic Norton dynamics'' provides an equivalent measure of the linear response, and demonstrated that this equivalence extends well beyond the linear response regime. In this article, our aim is to extend the results in~\cite{Blassel2023} to consider the generalisation of the stochastic Norton dynamics in DPD, important for a wide range of complex fluids and soft matter applications. Moreover, we will compare the performance of the stochastic Norton dynamics with NEMD in the DPD context, focusing on the computation of the mobility and the shear viscosity.

The rest of the article is organised as follows. In Section~\ref{sec:Mathematical_Formulations}, we describe the mathematical formulations of DPD and its variants with NEMD (i.e., DPD-NEMD) and the stochastic Norton dynamics (i.e., DPD-Norton), respectively. Section~\ref{sec:Numerical Methods} discusses the numerical methods used for the integration of the stochastic Norton dynamics in DPD. A variety of numerical experiments are performed in Section~\ref{sec:Numerical_Experiments} to compare all the approaches described in the article for the computation of the mobility and the shear viscosity. Our findings are summarised in Section~\ref{sec:Conclusions}.

\section{Mathematical formulations}
\label{sec:Mathematical_Formulations}

In this section, we introduce the mathematical formulations of DPD and its DPD-NEMD and DPD-Norton variants, respectively.

\subsection{DPD}
\label{subsec:DPD}

Originally updated in discrete timesteps, DPD was later reformulated by Espa\~nol and Warren~\cite{Espanol1995} as a proper statistical mechanics model written as a system of It\=o SDEs. Consider an $N$-particle system evolving in dimension $d$ with position $\q \in \mathbb{R}^{dN}$ and momentum $\p \in \mathbb{R}^{dN}$, the equations of motion of the DPD system (known as the ``reference dynamics''), in a vector form, are given by~\cite{Leimkuhler2016a}
\begin{subequations}\label{eq:DPD}
\begin{align}
  \dd \q &= \M^{-1}\p \dd t \, ,\\
  \dd \p &= -\nabla U(\q) \dd t - \gamma\GammaB(\q)\M^{-1}\p \dd t + \sigma\SigmaB(\q) \dd \WB \, ,
\end{align}
\end{subequations}
where $\M\in\mathbb{R}^{dN\times dN}$ is the diagonal mass matrix, $-\nabla U$ is the conservative force represented in terms of a potential energy function $U=U(\q)$ ($\nabla$ represents the gradient operator), $\gamma$ and $\sigma$ represent the dissipative and random strengths, respectively, and $\WB\in\mathbb{R}^{S \times 1}$ is a vector of $S=N(N-1)/2$ independent Wiener processes with mean zero and variance $\dd t$. Moreover, $\GammaB(\q) \in \mathbb{R}^{dN \times dN}$  and $\SigmaB(\q) \in \mathbb{R}^{dN \times S}$ are two position-dependent matrices that satisfy the following relation:
\begin{equation}\label{eq:FDR Matrices}
  \GammaB = \SigmaB\SigmaB^{\TT} \, ,
\end{equation}
where the matrix $\SigmaB$ can be written explicitly as, when $N \geq 5$,
%where the symmetric positive semi-definite matrix $\GammaB$ can be written explicitly as
%\begin{equation}
%  \GammaB(\mathbf{q})=\left(\begin{array}{cccc}
%    \sum\limits_{j\neq 1}\omega^{\D}(r_{1j})\e_{1j}\e_{1j}^{\TT} &  -\omega^{\D}(r_{12})\e_{12}\e_{12}^{\TT} & \cdots & -\omega^{\D}(r_{1N})\e_{1N}\e_{1N}^{\TT} \\
%    -\omega^{\D}(r_{21})\e_{21}\e_{21}^{\TT} & \sum\limits_{j\neq 2}\omega^{\D}(r_{2j})\e_{2j}\e_{2j}^{\TT} & \cdots & -\omega^{\D}(r_{2N})\e_{1N}\e_{1N}^{\TT} \\
%    \vdots & \vdots & \ddots & \vdots \\
%    -\omega^{\D}(r_{N1})\e_{N1}\e_{N1}^{\TT} & -\omega^{\D}(r_{N2})\e_{N2}\e_{N2}^{\TT} & \cdots & \sum\limits_{j\neq N}\omega^{\D}(r_{Nj})\e_{Nj}\e_{Nj}^{\TT}
%  \end{array}\right) \, ,
%\end{equation}
%
%\begin{equation}
%  \SigmaB(\mathbf{q})=\left(\begin{array}{cccccccc}
%    \omega^{\R}(r_{12})\e_{12} & \omega^{\R}(r_{13})\e_{13} & \cdots & \omega^{\R}(r_{1N})\e_{1N} & \mathbf{0} & \mathbf{0} & \cdots & \mathbf{0} \\
%    \omega^{\R}(r_{21})\e_{21} & \mathbf{0} & \cdots & \mathbf{0} & \omega^{\R}(r_{23})\e_{23} & \omega^{\R}(r_{24})\e_{24} & \cdots & \mathbf{0} \\
%    \mathbf{0} & \omega^{\R}(r_{31})\e_{31} & \cdots & \mathbf{0} & \omega^{\R}(r_{32})\e_{32} & \mathbf{0} & \cdots & \mathbf{0} \\
%    \vdots & \vdots & \ddots & \vdots & \vdots & \vdots & \iddots & \vdots \\
%    \mathbf{0} & \mathbf{0} & \cdots & \omega^{\R}(r_{N1})\e_{N1} & \mathbf{0} & \mathbf{0} & \cdots & \omega^{\R}(r_{N,N-1})\e_{N,N-1}
%  \end{array}\right) \, ,
%\end{equation}
\begin{equation}
  \SigmaB(\mathbf{q})=\left(\begin{array}{cccccccc}
    \omega_{12}^{\R}\e_{12} & \omega_{13}^{\R}\e_{13} & \cdots & \omega_{1N}^{\R}\e_{1N} & \mathbf{0} & \mathbf{0} & \cdots & \mathbf{0} \\
    \omega_{21}^{\R}\e_{21} & \mathbf{0} & \cdots & \mathbf{0} & \omega_{23}^{\R}\e_{23} & \omega_{24}^{\R}\e_{24} & \cdots & \mathbf{0} \\
    \mathbf{0} & \omega_{31}^{\R}\e_{31} & \cdots & \mathbf{0} & \omega_{32}^{\R}\e_{32} & \mathbf{0} & \cdots & \mathbf{0} \\
    \vdots & \vdots & \ddots & \vdots & \vdots & \vdots & \iddots & \vdots \\
    \mathbf{0} & \mathbf{0} & \cdots & \omega_{N1}^{\R}\e_{N1} & \mathbf{0} & \mathbf{0} & \cdots & \omega_{N,N-1}^{\R}\e_{N,N-1}
  \end{array}\right) \, ,
\end{equation}
where 
%$\omega^{\D}\left(r_{ij}\right)$ 
$\omega_{ij}^{\R} = \omega^{\R}\left(r_{ij}\right)$ is a weight function defined in the DPD system~\cite{Leimkuhler2015,Leimkuhler2016a}
%, \textcolor{red}{$\mathbf{E}_{ij} = \e_{ij}\e_{ij}^{\TT} \in \mathbb{R}^{d \times d}$ is a matrix,} 
and $\mathbf{e}_{ij} = \left(\q_{i}-\q_{j}\right)/r_{ij}$ is the unit vector pointing from particle $j$ to particle $i$, with $r_{ij} = \left|\q_{ij}\right| = \left|\q_{i}-\q_{j}\right|$ being the distance between the two particles. 

In addition to~\eqref{eq:FDR Matrices}, the following relation needs to be satisfied
\begin{equation}\label{eq:FDR}
  \sigma^{2} = 2\gamma\kB T \, ,
\end{equation}
where $\kB$ and $T$ are the Boltzmann constant and the equilibrium temperature, respectively, in the DPD system~\eqref{eq:DPD} in order to preserve the momentum-constrained canonical ensemble with density 
\begin{equation}
  \rho_{\beta}(\q,\p) = {Z}^{-1} \exp(-\beta H(\q,\p)) \times \delta\!\left[ \sum_i p^{x}_{i}- \pi_x \right] \delta\!\left[\sum_i p^{y}_{i} - \pi_y \right] \delta\!\left[\sum_i p^{z}_{i} - \pi_z \right] \, ,
\end{equation}
where $Z$ is a suitable normalising constant (i.e., the partition function), $\beta^{-1}=\kB T$, $H(\q,\p) = \p^{\TT}\M^{-1}\p/2 + U(\q)$ represents the system Hamiltonian, and $\boldsymbol{\pi}=\left(\pi_x,\pi_y,\pi_z\right)$ is the linear momentum vector. It is worth mentioning that the ergodicity of the DPD system has only been established in the case of high particle density in one dimension by Shardlow and Yan~\cite{Shardlow2006}; extending the result to higher dimensions with more general conditions is highly nontrivial.  

\subsection{DPD-NEMD}
\label{subsec:DPD-NEMD}

In standard NEMD approaches, we consider adding an external nongradient force, in the form of a smooth vector field $\F \in \mathbb{R}^{dN}$, in the drift of a ``reference dynamics'' in equilibrium, leading to a ``perturbed dynamics''. As in~\cite{Blassel2023}, we restrict our attention in this article to position-dependent external force, i.e., $\F = \F(\q)$, in the NEMD method applied to the DPD system~\eqref{eq:DPD}. In this case, the equations of motion of the DPD-NEMD dynamics are given by
\begin{subequations}\label{eq:DPD NEMD}
\begin{align}
  \dd \q &= \M^{-1}\p \dd t \, ,\\
  \dd \p &= [-\nabla U(\q)+\eta\F(\q)] \dd t - \gamma\GammaB(\q)\M^{-1}\p \dd t + \sigma\SigmaB(\q) \dd \WB \, ,
\end{align}
\end{subequations}
where $\eta \in \mathbb{R}$ represents the strength of the force. In the case of $\eta=0$, the perturbed dynamics~\eqref{eq:DPD NEMD} reduces to the reference dynamics~\eqref{eq:DPD}. As in~\cite{Blassel2023}, we assume that, if $|\eta|$ is small enough, the perturbed dynamics~\eqref{eq:DPD NEMD} is ergodic and admits a unique invariant probability measure. The corresponding expectation associated with the perturbed invariant probability measure is denoted by $\E_\eta$. Based on the linear response theory~\cite{Hairer2010,Joubaud2012}, for a given response observable $R \in \mathbb{R}$ such that
\begin{equation}\label{eq:NEMD centering condition}
  \E_0[R] = 0
\end{equation}
(i.e., zero expectation associated with the invariant probability measure of the reference dynamics~\eqref{eq:DPD}), the associated transport coefficient $\alpha$ can be defined as
\begin{equation}\label{eq:NEMD transport}
  \alpha = \lim \limits_{\eta\rightarrow 0} \frac{\E_\eta[R]}{\eta} \, ,
\end{equation}
provided that it is well defined. The definition above suggests that one can first compute ergodic averages of $R$ over trajectories of the perturbed dynamics~\eqref{eq:DPD NEMD}, and then estimate the linear relation between $R$ and $\eta$ for one or several values of $\eta$ in the linear response regime. We can further write down the finite difference estimator for the limit~\eqref{eq:NEMD transport} as the following ergodic average:
\begin{equation}\label{eq:NEMD transport estimator}
  \widehat{\alpha}_{T,\eta} = \frac{1}{\eta T} \int_{0}^{T} R\left(\q_t,\p_t\right) \, \dd t \, .
\end{equation}
%However, the challenge associated with the NEMD method is that the estimator~\eqref{eq:NEMD transport estimator} suffers from large statistical errors when $|\eta|$ is small. More precisely, we can show that the asymptotic variance of the estimator~\eqref{eq:NEMD transport estimator} scales as $\eta^{-2}$ as $\eta$ approaches zero~\cite{Lelievre2016,Blassel2023}.

%Although there exist techniques to extend the linear regime~\cite{Spacek2023}, in many cases the linear regimes are limited, resulting in large asymptotic variances. Furthermore, we have
%\begin{equation}\label{eq:NEMD var}
%  \sigma^2_{\alpha,\eta} = \frac{2}{\eta^2} \Var_\eta(R) \Theta_\eta(R) \, ,
%\end{equation}
%where $\Var_\eta$ denotes the variance with respect to $\E_\eta$ and the correlation time $\Theta_\eta(R)$ is given by
%\begin{equation}\label{eq:NEMD auto}
%  \Theta_\eta(R) = \int_0^\infty \frac{\E_\eta\left[\Pi_{\eta}R(\X^\eta_t) \Pi_{\eta}R(\X^\eta_0) \right]}{\E_\eta\left[(\Pi_{\eta}R)^2\right]} \, \dd t \, ,
%\end{equation}
%where $\Pi_{\eta}$ is the centering operator given by $\Pi_{\eta}\varphi = \varphi - \E_\eta[\varphi]$.

\subsection{DPD-Norton}
\label{subsec:DPD-Norton}

While standard NEMD approaches, as shown in the previous subsection, fix the forcing and measure the average induced flux in the system driven out of equilibrium, an alternative approach, the stochastic Norton dynamics, instead fixes the value of the flux and measures the average magnitude of the forcing needed to induce it~\cite{Blassel2023}. In the stochastic Norton dynamics approach, we only consider the response observable $R$ in the following specific form, as in~\cite{Blassel2023}:
\begin{equation}\label{eq:response}
  R(\q,\p) = \G(\q) \cdot \p \, ,
\end{equation}
where $\G \in \mathbb{R}^{dN}$ is a position-dependent vector function. When applied to the DPD system~\eqref{eq:DPD}, the equations of motion of the DPD-Norton dynamics are given by
\begin{subequations}\label{eq:DPD Norton}
\begin{align}
  \dd \q_t &= \M^{-1}\p_t \dd t \, ,\\
  \dd \p_t &= -\nabla U\left(\q_t\right) \dd t - \gamma\GammaB\left(\q_t\right)\M^{-1}\p_t \dd t + \sigma\SigmaB\left(\q_t\right) \dd \WB_t +\F\left(\q_t\right) \dd \Lambda_{t} \, ,\\
  R\left(\q_t,\p_t\right) &= R\left(\q_0,\p_0\right) = r \, ,
\end{align}
\end{subequations}
where $r \in \mathbb{R}$ represents the strength of the response, and $\Lambda_{t}$ is a stochastic process required to constrain the response. Moreover, $\Lambda_{t}$ can be defined as an It\^{o} process and decomposed as follows:
\begin{equation}\label{eq:Ito_forcing}
  \Lambda_{t} = \Lambda_{0} + \int_0^t \lambda\left(\q_s,\p_s\right) \, \dd s + \widetilde{\Lambda}_{t} \, , \quad \widetilde{\Lambda}_{t} = \int_0^t \widetilde{\boldsymbol{\lambda}}\left(\q_s,\p_s\right) \, \dd \WB_s \, ,
\end{equation}
where $\lambda\in\mathbb{R}$ and $\widetilde{\boldsymbol{\lambda}} \in \mathbb{R}^{1 \times S}$ are two functions. In what follows, unless necessary, the subscripts associated with $\q$, $\p$, $\Lambda$, and $\WB$ will be dropped for notational simplicity. Note that, neglecting the zero-mean contribution of $\widetilde{\boldsymbol{\lambda}}$, the average forcing in the stochastic Norton dynamics can then be defined as the expectation of $\lambda$, also known as the forcing observable. Based on the form of the response observable $R$ in~\eqref{eq:response}, we can write down the expression of the forcing observable $\lambda$ associated with the DPD-Norton dynamics~\eqref{eq:DPD Norton}:
\begin{equation}\label{eq:DPD Norton lambda}
  \lambda(\q,\p) = \frac{1}{\F(\q)\cdot\G(\q)} \left( \G(\q) \cdot \left[ \nabla U(\q) + \gamma\GammaB(\q)\M^{-1}\p \right] - \nabla\G(\q)\p \cdot \M^{-1}\p \right) \, .
\end{equation}
Note that the controllability condition $\G(\q) \cdot \F(\q) \neq0$ needs to be satisfied. As in~\cite{Blassel2023}, we assume the existence and uniqueness of the invariant steady-state probability measure for the DPD-Norton dynamics~\eqref{eq:DPD Norton}, whose expectation is denoted by $\E^*_r$. We further assume that
\begin{equation}\label{eq:Norton centering condition}
  \E^*_0[\lambda] = 0 \, ,
\end{equation}
which is the Norton counterpart of the centering condition~\eqref{eq:NEMD centering condition} for the observable in DPD-NEMD simulations. When~\eqref{eq:Norton centering condition} holds, the transport coefficient for the DPD-Norton dynamics~\eqref{eq:DPD Norton} can be defined by analogy with~\eqref{eq:NEMD transport} as
\begin{equation}\label{eq:Norton transport}
  \alpha^* = \lim \limits_{r\rightarrow 0} \frac{r}{\E^*_r[\lambda]} \, ,
\end{equation}
provided that the limit exists. Assuming the DPD-Norton dynamics~\eqref{eq:DPD Norton} is ergodic with respect to the steady-state, a natural estimator of the transport coefficient~\eqref{eq:Norton transport} can be constructed by replacing ensemble averages by trajectory averages:
\begin{equation}\label{eq:Norton transport estimator}
  \widehat{\alpha}^*_{T,r} = \frac{rT}{\int_{0}^{T} \lambda\left(\q_t,\p_t\right) \, \dd t} \, ,
\end{equation}
which is similar to what is done in DPD-NEMD simulations~\eqref{eq:NEMD transport estimator}.

%Similar to the analyses in NEMD in the previous subsection, we can show that the asymptotic variance of the estimator~\eqref{eq:Norton transport estimator} scales as $r^{-2}$ as $r$ approaches zero~\cite{Blassel2023}. More precisely, we have
%\begin{equation}\label{eq:Norton var}
%  \sigma^2_{\alpha^*, r} = \frac{2r^2}{(\E^*_r[\lambda])^4} \Var_r^*(\lambda) \Theta_r^*(\lambda) = \frac{2(\alpha^*)^4}{r^2} \Var_r^*(\lambda) \Theta_r^*(\lambda) + O \left( \frac{1}{r} \right) \, ,
%\end{equation}
%where $\Var_r^*$ denotes the variance with respect to $\E^*_r$ and the correlation time $\Theta_r^*(\lambda)$ is defined similarly to the NEMD case~\eqref{eq:NEMD auto} by
%\begin{equation}\label{eq:Norton auto}
%  \Theta^*_r(\lambda) = \int_0^\infty \frac{\E^*_r\left[ \Pi^*_r \lambda(\Y^r_t) \Pi^*_r \lambda(\Y^r_0) \right]}{\E^*_r\left[ (\Pi^*_r \lambda)^2 \right]} \dd t \, ,
%\end{equation}
%where $\Pi^*_r$ is the centering operator with respect to $\E^*_r$.

\section{Numerical methods}
\label{sec:Numerical Methods}

A great deal of effort has been devoted to developing accurate and efficient numerical methods to numerically discretise DPD and related systems. It has been demonstrated that splitting methods often outperform alternative schemes in various settings~\cite{Shardlow2003,Leimkuhler2015,Leimkuhler2016a,Shang2017,Shang2018,Shang2020}. In this section, we will apply splitting methods in both the DPD-NEMD and DPD-Norton dynamics.

\subsection{DPD-NEMD}
\label{subsec:DPD-NEMD}

We decompose the vector field of the DPD-NEMD system~\eqref{eq:DPD NEMD} into three pieces, labeled as $\Ar$, $\Br$, and $\Or$, respectively:
\begin{equation}
  \dd \left[ \begin{array}{c} \q \\ \p \end{array} \right] = \underbrace{\left[ \begin{array}{c} \M^{-1}\p \\ \mathbf{0} \end{array} \right] \dd t}_\Ar + \underbrace{\left[ \begin{array}{c} \mathbf{0} \\ -\nabla U(\q) + \eta\F(\q)\end{array} \right] \dd t }_\Br + \underbrace{\left[ \begin{array}{c} \mathbf{0} \\ -\gamma\GammaB(\q)\M^{-1}\p \dd t + \sigma\SigmaB(\q) \dd \WB \end{array} \right]}_\Or \, .
\end{equation}
We can write down the generator of each piece as follows:
\begin{subequations}
\begin{align}
  \Lg_\Ar &= \M^{-1}\p \cdot \nabla_\q \, ,\\
  \Lg_\Br &= -\nabla U(\q) \cdot \nabla_\p + \eta\F(\q) \cdot \nabla_\p \, ,\\
  \Lg_\Or &= -\gamma\GammaB(\q)\M^{-1}\p \cdot \nabla_\p + \frac{\sigma^2}{2}\SigmaB(\q)\left[\SigmaB(\q)\right]^{\TT}:\nabla^2_\p \, ,
\end{align}
\end{subequations}
where $\nabla^2$ is the Hessian matrix and $:$ represents the Frobenius inner product defined as $\A:\B=\Tr\left(\A^\TT\B\right)$. The generator of the DPD-NEMD system can be written as
\begin{equation}
  \Lg = \Lg_\Ar + \Lg_\Br + \Lg_\Or \, .
\end{equation}
The elementary dynamics associated with $\Lg_\Ar$ is given by
\begin{subequations}\label{eq:DPD NEMD A}
\begin{align}
  \dd \q &= \M^{-1}\p \dd t \, ,\\
  \dd \p &= \mathbf{0} \, ,
\end{align}
\end{subequations}
while the elementary dynamics associated with $\Lg_\Br$ is given by
\begin{subequations}\label{eq:DPD NEMD B}
\begin{align}
  \dd \q &= \mathbf{0} \, ,\\
  \dd \p &= [-\nabla U(\q)+\eta\F(\q)] \dd t \, .
\end{align}
\end{subequations}
Both elementary dynamics~\eqref{eq:DPD NEMD A}--\eqref{eq:DPD NEMD B} can be solved exactly. The elementary dynamics associated with $\Lg_\Or$ is an Ornstein--Uhlenbeck process on the momentum
\begin{subequations}\label{eq:DPD NEMD O}
\begin{align}
  \dd \q &= \mathbf{0} \, ,\\
  \dd \p &= - \gamma\GammaB(\q)\M^{-1}\p \dd t + \sigma\SigmaB(\q) \dd \WB \, ,
\end{align}
\end{subequations}
and can also be solved exactly based on each interacting pair (see more details in~\cite{Leimkuhler2015,Leimkuhler2016,Shang2020}). Moreover, the flow map (or phase space propagator) of the DPD-NEMD system is given by the shorthand notation
\begin{equation}
  \mathcal{F}_{t} = \exp \left( t \mathcal{L} \right) \, ,
\end{equation}
where the exponential map is used to formally denote the solution operator. A number of approximations of $\mathcal{F}_{t}$ can be obtained as products (taken in different arrangements) of exponentials of the splitting terms. However, it turns out that different splittings (other than ``A'', ``B'', and ``O'' in the current article, e.g., the stochastic position Verlet and stochastic velocity Verlet methods in~\cite{Melchionna2007} where the system is split into two elementary dynamics) and/or combinations give dramatically different performance in practice~\cite{Shardlow2003,Leimkuhler2015,Leimkuhler2016a,Shang2017,Shang2020}. In particular, it has been demonstrated in various numerical experiments, on both equilibrium and
transport properties that the so-called ``ABOBA'' splitting method~\cite{Shang2020} outperforms popular alternative schemes in the literature. Therefore, we will apply the ABOBA method in both NEMD and stochastic Norton dynamics approaches in DPD. To this end, the phase space propagation of the ABOBA method can be written as
\begin{equation}\label{eq:Propagator DPD NEMD}
  e^{\Delta t\Lg_\mathrm{ABOBA}} = e^{\frac{\Delta t}{2}\Lg_\Ar} e^{\frac{\Delta t}{2}\Lg_\Br} e^{\Delta t\Lg_\Or} e^{\frac{\Delta t}{2}\Lg_\Br} e^{\frac{\Delta t}{2}\Lg_\Ar} \, ,
\end{equation}
%\begin{equation}\label{eq:Propagator DPD NEMD}
%  \exp\left(\Delta t\Lg_\mathrm{ABOBA}\right) = \exp\left(\frac{\Delta t}{2}\Lg_\Ar\right) \exp\left(\frac{\Delta t}{2}\Lg_\Br\right) \exp\left(\Delta t\Lg_\Or\right) \exp\left(\frac{\Delta t}{2}\Lg_\Br\right) \exp\left(\frac{\Delta t}{2}\Lg_\Ar\right) \, ,
%\end{equation}
where $\exp\left(\Delta t\mathcal{L}_f\right)$ denotes the phase space propagator associated with the corresponding vector field $f$. Note that the steplengths associated with various operations are uniform and span the interval $\Delta t$. Therefore, each of the A and B steps in~\eqref{eq:Propagator DPD NEMD} is taken with a steplength of $\Delta t/2$, while a steplength of $\Delta t$ is used in the O step. The detailed integration steps of the ABOBA method for the DPD-NEMD system can be written as follows:

\noindent \textbf{Step 1}: for all particles,
\begin{align}
  \q^{n+1/2} &= \q^{n} + (\Delta t/2) \M^{-1}\p^{n}\, ,\\
  \p^{n+1/3} &= \p^{n} - (\Delta t/2) \nabla U\left(\q^{n+1/2}\right) + (\Delta t/2) \eta\F\left(\q^{n+1/2}\right)\, .
\end{align}

\noindent \textbf{Step 2}: for each interacting pair within cutoff radius ($r_{ij}<r_{\mathrm{c}}$), where $r_{\mathrm{c}}$ is the cutoff radius, in a successive manner,
\begin{align}
  \p_{i}^{n+2/3} &= \p_{i}^{n+1/3} + m_{ij} \left[ \Delta v_{ij}^{\mathrm{D}}\left(\q^{n+1/2},\p^{n+1/3}\right) + \Delta v_{ij}^{\mathrm{R}}\left(\q^{n+1/2}\right) \right] \e_{ij}^{n+1/2} \, ,\\
  \p_{j}^{n+2/3} &= \p_{j}^{n+1/3} - m_{ij} \left[ \Delta v_{ij}^{\mathrm{D}}\left(\q^{n+1/2},\p^{n+1/3}\right) + \Delta v_{ij}^{\mathrm{R}}\left(\q^{n+1/2}\right) \right] \e_{ij}^{n+1/2} \, ,
\end{align}
where $m_{ij}=m_{i}m_{j}/\left(m_{i}+m_{j}\right)$, with $m_{i}$ being the mass of the particle $i$, is the ``reduced mass'' and
\begin{equation}
  \Delta v_{ij}^{\mathrm{D}}(\q,\p) = \left[ \mathbf{e}_{ij} \cdot \mathbf{v}_{ij} \right] \left( e^{-\tau\left(r_{ij}\right)\Delta t} - 1 \right) \, , \quad \Delta v_{ij}^{\mathrm{R}}(\q) = \frac{\sigma \omega^{\mathrm{R}}\left(r_{ij}\right)}{m_{ij}} \sqrt{ \frac{1-e^{-2\tau\left(r_{ij}\right) \Delta t}}{2\tau\left(r_{ij}\right)}}\mathrm{R}^{n}_{ij} \, ,
\end{equation}
where $\mathbf{v}_{ij} = \mathbf{p}_{i}/m_{i}-\mathbf{p}_{j}/m_{j}$ is the relative velocity, $\tau\left(r_{ij}\right) = \gamma \omega^{\mathrm{D}}\left(r_{ij}\right) / m_{ij}$, $\mathrm{R}_{ij}$ is a normally distributed variable with zero mean and unit variance. $\omega^{\D}\left(r_{ij}\right)$ is another weight function~\cite{Leimkuhler2015,Leimkuhler2016a} that satisfies the relation of $\omega^{\mathrm{D}}\left(r_{ij}\right)=\left[\omega^{\mathrm{R}}\left(r_{ij}\right)\right]^{2}$. Note that one of the two weight functions can be arbitrarily chosen, for instance, a popular choice of $\omega^{\mathrm{R}}\left(r_{ij}\right)$ that is also used in this article is given by
\begin{equation}\label{eq:omega_R}
  \omega^{\mathrm{R}}\left(r_{ij}\right)=
  \begin{cases}
  1-r_{ij}/r_{\mathrm{c}} \, , & r_{ij}<r_{\mathrm{c}} \, ,\\
  \quad \quad 0 \, , & r_{ij}\geq r_{\mathrm{c}} \, .
  \end{cases}
\end{equation}

\noindent \textbf{Step 3}: for all particles,
\begin{align}
  \p^{n+1} &= \p^{n+2/3} - (\Delta t/2) \nabla U\left(\q^{n+1/2}\right) + (\Delta t/2) \eta\F\left(\q^{n+1/2}\right) \, ,\\
  \q^{n+1} &= \q^{n+1/2} + (\Delta t/2) \M^{-1}\p^{n+1} \, .
\end{align}

\subsection{DPD-Norton}
\label{subsec:DPD-Norton}

We can similarly decompose the vector field of the DPD-Norton system~\eqref{eq:DPD Norton} into three pieces, again labeled as $\Ar$, $\Br$, and $\Or$, respectively:
\begin{equation}\label{eq:DPD Norton Splitting}
\begin{aligned}
  \dd \left[ \begin{array}{c} \q \\ \p \end{array} \right] = &\, \underbrace{\left[ \begin{array}{c} \M^{-1}\p\dd t \\ \F(\q)\dd\Lambda^{\Ar} \end{array} \right] }_\Ar + \underbrace{\left[ \begin{array}{c} \mathbf{0} \\ -\nabla U(\q)\dd t + \F(\q)\dd\Lambda^{\Br} \end{array} \right]  }_\Br \\
  &+ \underbrace{\left[ \begin{array}{c} \mathbf{0} \\ -\gamma\GammaB(\q)\M^{-1}\p \dd t + \sigma\SigmaB(\q) \dd \WB  +\F(\q)\dd\Lambda^{\Or} \end{array} \right]}_\Or \, ,
\end{aligned}
\end{equation}
where the stochastic process $\Lambda$ are divided into three parts corresponding to each piece, $\Lambda^{\Ar}$, $\Lambda^{\Br}$, and $\Lambda^{\Or}$, respectively. As demonstrated in~\cite{Blassel2023}, the expressions of the function $\lambda$ and the martingale $\widetilde{\Lambda}$ can be made precise, thereby allowing us to write down the DPD-Norton dynamics without explicitly referencing to the forcing. We can write down the generator of each piece as follows:
\begin{subequations}
\begin{align}
  \mathfrak{L}_\Ar &= \M^{-1}\p\cdot\nabla_\q - \frac{\nabla\G(\q)\p\cdot\M^{-1}\p}{\F(\q)\cdot\G(\q)} \F(\q) \cdot \nabla_\p \, ,\\
  \mathfrak{L}_\Br &= -\PP_{\F,\G}(\q)\nabla U(\q) \cdot \nabla_\p \, ,\\
  \mathfrak{L}_\Or &= -\gamma\PP_{\F,\G}(\q)\GammaB(\q)\M^{-1}\p \cdot \nabla_\p + \frac{\sigma^2}{2} \PP_{\F,\G}(\q)\SigmaB(\q) \left[\PP_{\F,\G}(\q)\SigmaB(\q)\right]^{\TT}:\nabla^2_\p \, ,
\end{align}
\end{subequations}
where the nonorthogonal projector-valued map $\PP_{\F,\G}(\q)$ is given by
\begin{equation}
  \PP_{\F,\G}(\q) = \I - \frac{\F(\q) \otimes \G(\q)}{\F(\q) \cdot \G(\q)} \, ,  
\end{equation}
where $\otimes$ represents the Kronecker product and $\I$ is an identity matrix. The generator of the DPD-Norton system can then be written as
\begin{equation}
  \mathfrak{L} = \mathfrak{L}_\Ar + \mathfrak{L}_\Br + \mathfrak{L}_\Or \, .
\end{equation}
It is worth mentioning that the decomposition~\eqref{eq:DPD Norton Splitting} is motivated by the fact that each elementary dynamics individually preserves the response observable~\eqref{eq:response}, i.e., $\mathfrak{L}_\Ar R = \mathfrak{L}_\Br R = \mathfrak{L}_\Or R =0$. Therefore, the solutions and approximation to the elementary dynamics are designed in such a way that each of them preserves the response.

The elementary dynamics associated with $\mathfrak{L}_\Br$ is given by
\begin{subequations}\label{eq:DPD Norton B}
\begin{align}
  \dd \q &= \mathbf{0} \, ,\\
  \dd \p &= -\PP_{\F,\G}(\q)\nabla U(\q) \dd t \, ,
\end{align}
\end{subequations}
while the elementary dynamics associated with $\mathfrak{L}_\Or$ is given by
\begin{subequations}\label{eq:DPD Norton O}
\begin{align}
  \dd \q &= \mathbf{0} \, ,\\
  \dd \p &= \PP_{\F,\G}(\q) \left[ -\gamma\GammaB(\q)\M^{-1}\p \dd t + \sigma\SigmaB(\q) \dd \WB \right] \, .
\end{align}
\end{subequations}
Similar to the cases for the DPD-NEMD system in the previous subsection, both elementary dynamics~\eqref{eq:DPD Norton B}--\eqref{eq:DPD Norton O} can be solved exactly. However, the elementary dynamics associated with $\mathfrak{L}_\Ar$
\begin{subequations}\label{eq:DPD Norton A}
\begin{align}
  \dd \q &= \M^{-1}\p \dd t \, ,\\
  \dd \p &= -\frac{\nabla\G(\q)\p\cdot\M^{-1}\p}{\F(\q) \cdot \G(\q)} \F(\q) \dd t \, ,
\end{align}
\end{subequations}
is not exactly solvable, and has to be numerically approximated. 

In describing the splitting method associated with the DPD-Norton dynamics~\eqref{eq:DPD Norton Splitting}, we introduce a discrete flow, $\Phi_{\Delta t,r}$, associated with each elementary dynamics. More specifically, the discrete flow associated with the elementary A-dynamics~\eqref{eq:DPD Norton A} is given by
\begin{equation}\label{eq:DPD Norton Phi A}
  \Phi^\Ar_{\Delta t, r}(\q,\p,\ell) = \left( \q + \Delta t \M^{-1} \p, \p + \xi_{\Delta t, r}^\Ar(\q,\p) \F\left(\q+\Delta t\M^{-1}\p\right), \ell + \xi_{\Delta t, r}^\Ar(\q,\p) \right) \, ,
\end{equation}
where $\xi^\Ar_{\Delta t, r} \in \mathbb{R}$ is a Lagrange multiplier to ensure the response will be preserved at a constant $r$ for each elementary dynamics, and $\ell \in \mathbb{R}$ is an auxiliary variable  to accumulate the bounded-variation component of Lagrange multipliers during the integration step (see more details in~\cite{Blassel2023}). When the response observable $R$ is of a specific form as in~\eqref{eq:response}, we can explicitly solve
%, $\G\left(\q + \Delta t \M^{-1} \p\right) \cdot \left( \p + \xi_{\Delta t, r}^\Ar(\q,\p) \F\left(\q+\Delta t\M^{-1}\p\right) \right) = r$, 
\begin{equation}\label{eq:DPD Norton Constant Flux A}
  \G\left(\q + \Delta t \M^{-1} \p\right) \cdot \left( \p + \xi_{\Delta t, r}^\Ar(\q,\p) \F\left(\q+\Delta t\M^{-1}\p\right) \right) = r
\end{equation}
for the Lagrange multiplier $\xi^\Ar_{\Delta t, r}(\q,\p)$:
\begin{equation}\label{eq:DPD Norton xi A}
  \xi^\Ar_{\Delta t, r}(\q,\p) = \frac{r-\G\left(\q + \Delta t \M^{-1} \p\right) \cdot \p}{\F\left(\q + \Delta t \M^{-1} \p\right) \cdot \G\left(\q + \Delta t \M^{-1} \p\right)} \, .
\end{equation}
Similarly, the discrete flow associated with the elementary B-dynamics~\eqref{eq:DPD Norton B} is given by
\begin{equation}\label{eq:DPD Norton Phi B}
  \Phi^\Br_{\Delta t, r}(\q,\p,\ell) = \left( \q, \p - \Delta t \nabla U(\q) + \xi_{\Delta t, r}^\Br(\q,\p) \F(\q), \ell+\xi_{\Delta t, r}^\Br(\q,\p) \right) \, .
\end{equation}
Again, for a specific form of the response observable $R$ as in~\eqref{eq:response}, we can explicitly solve
%, $\G(\q) \cdot \left( \p - \Delta t \nabla U(\q) + \xi_{\Delta t, r}^\Br(\q,\p) \F(\q) \right) = r$, 
\begin{equation}\label{eq:DPD Norton Constant Flux B}
  \G(\q) \cdot \left( \p - \Delta t \nabla U(\q) + \xi_{\Delta t, r}^\Br(\q,\p) \F(\q) \right) = r
\end{equation}
for the Lagrange multiplier $\xi^\Br_{\Delta t, r}(\q,\p)$:
\begin{equation}\label{eq:DPD Norton xi B}
  \xi^\Br_{\Delta t, r}(\q,\p) = \frac{r-\G(\q) \cdot \left( \p - \Delta t \nabla U(\q) \right)}{\F(\q) \cdot \G(\q)} \, ,
\end{equation}
in which case,~\eqref{eq:DPD Norton Phi B} coincides with the analytical integration of the elementary B-dynamics~\eqref{eq:DPD Norton B} over one time step $\Delta t$ when the constant flux condition~\eqref{eq:DPD Norton Constant Flux B} is satisfied. Note that, unlike the other two elementary dynamics (i.e.,~\eqref{eq:DPD Norton B} and ~\eqref{eq:DPD Norton A}), the discrete flow associated with the elementary O-dynamics~\eqref{eq:DPD Norton O} is stochastic, and is given by
\begin{equation}\label{eq:DPD Norton Phi O}
  \widehat{\Phi}^\Or_{\Delta t, r}(\q,\p,\Delta \p,\ell) = \widehat{\Phi}^{\Or_\xi}_{\Delta t, r}\circ\widehat{\Phi}^{\Or_{N-1,N}}_{\Delta t, r}\circ\cdots\circ\widehat{\Phi}^{\Or_{1,3}}_{\Delta t, r}\circ\widehat{\Phi}^{\Or_{1,2}}_{\Delta t, r}(\q,\p,\mathbf{0},0) \, ,
\end{equation}
where 
\begin{subequations}
\begin{align}
  \widehat{\Phi}^{\Or_{i,j}}_{\Delta t, r}\left(\q,\p,\Delta \p,\ell|R_{ij}\right) &= \left( \q, \p + m_{ij} \left[ \Delta v_{ij}^{\D}(\q,\p) + \Delta v_{ij}^{\R}(\q) \right] \widehat{\e}_{ij}, \Delta \p + m_{ij} \Delta v_{ij}^{\D}(\q,\p) \widehat{\e}_{ij}, \ell \right) \, ,\\
  \widehat{\Phi}^{\Or_\xi}_{\Delta t, r}\left(\q,\p,\Delta \p,\ell|R_{ij}\right) &= \left( \q, \p + \widehat{\xi}_{\Delta t, r}^\Or(\q,\p)\F(\q), \Delta \p, \ell -\frac{\G(\q)\cdot\Delta\p}{\F(\q)\cdot\G(\q)} \right) \, ,
\end{align}
\end{subequations}
with $\widehat{\e}_{ij}$ being defined as
\begin{equation}
  \widehat{\e}_{ij} = \left( \mathbf{0}, \dots, \mathbf{0}, \underbrace{\e_{ij}^\TT}_{d(i-1)+1,\dots,di}, \mathbf{0}, \dots, \mathbf{0}, \underbrace{-\e_{ij}^\TT}_{d(j-1)+1,\dots,dj}, \mathbf{0}, \dots, 0 \right)^\TT \in \mathbb{R}^{dN} \, .
\end{equation}
For a specific form of the response observable $R$ as in~\eqref{eq:response}, we can again explicitly solve
%, $\G(\q) \cdot \left( \p + m_{ij} \left[ \Delta v_{ij}^{\D}(\q,\p) + \Delta v_{ij}^{\R}(\q) \right] \widehat{\e}_{ij} + \widehat{\xi}_{\Delta t, r}^\Or(\q,\p)\F(\q) \right) = r$,
\begin{equation}\label{eq:DPD Norton Constant Flux O}
  \G(\q) \cdot \left( \p + \sum_{i} \sum_{j>i} m_{ij} \left[ \Delta v_{ij}^{\D}(\q,\p) + \Delta v_{ij}^{\R}(\q) \right] \widehat{\e}_{ij} + \widehat{\xi}_{\Delta t, r}^\Or(\q,\p)\F(\q) \right) = r
\end{equation}
for the Lagrange multiplier $\widehat{\xi}^\Or_{\Delta t, r}(\q,\p)$:
\begin{equation}\label{eq:DPD Norton xi O}
  \widehat{\xi}^\Or_{\Delta t, r}(\q,\p) = \frac{r - \G(\q) \cdot \left( \p + \sum_{i} \sum_{j>i} m_{ij} \left[ \Delta v_{ij}^{\mathrm{D}}(\q,\p) + \Delta v_{ij}^{\mathrm{R}}(\q) \right] \widehat{\e}_{ij} \right)}{\F(\q)\cdot\G(\q)} \, ,
\end{equation}
in which case,~\eqref{eq:DPD Norton Phi O} is expected to be a second order approximation of the analytical integration of the elementary O-dynamics~\eqref{eq:DPD Norton O} over one time step $\Delta t$ when the constant flux condition~\eqref{eq:DPD Norton Constant Flux O} is satisfied~\cite{Shang2020}. Since the contribution of the random noise $R_{ij}$ to the Lagrange multiplier $\widehat{\xi}^\Or_{\Delta t, r}(\q,\p)$ is a centered Gaussian, we can remove it entirely to only accumulate the nonmartingale component in $\ell$, which is equivalent to a variance reduction technique~\cite{Blassel2023}. 
We will again apply the ABOBA method, whose update rule can be written as
\begin{equation}\label{eq:Propagator DPD Norton}
  \left(\q^{n+1},\p^{n+1},\ell^{n+1}\right) = \Phi^\Ar_{\Delta t/2, r} \circ \Phi^\Br_{\Delta t/2, r} \circ \widehat{\Phi}^\Or_{\Delta t, r}\left(\cdot|R_{ij}\right) \circ \Phi^\Br_{\Delta t/2, r} \circ \Phi^\Ar_{\Delta t/2, r}\left(\q^{n},\p^{n},\ell^{n}\right) \, .
\end{equation}
%It is worth mentioning that the decomposition~\eqref{eq:DPD Norton Splitting} is motivated by the fact that each elementary dynamics individually preserves the response observable~\eqref{eq:response}, i.e., $\mathfrak{L}_\Ar R = \mathfrak{L}_\Br R = \mathfrak{L}_\Or R =0$. Therefore, the solutions and approximation to the elementary dynamics are designed in such a way that each of them preserves the response. 
%To this end, we introduce a Lagrange multiplier $\xi \in \mathbb{R}$ to ensure the response will be preserved at a constant $r$ for each elementary dynamics, and an auxiliary variable $\ell \in \mathbb{R}$ to accumulate the bounded-variation component of Lagrange multipliers during the integration step (see more details in~\cite{Blassel2023}). 

The forcing observable $\lambda$ can then be estimated by averaging the auxiliary variable $\ell$ in each increment. The detailed integration steps of the ABOBA method for the DPD-Norton system can be written as follows:

\noindent \textbf{Step 1}: starting from $\ell^{n}=0$, for all particles,
\begin{align}
  \q^{n+1/2} &= \q^n + (\Delta t/2)\M^{-1}\p^n \, ,\\
  \p^{n+1/5} &= \p^n + \xi^\Ar_{\Delta t/2, r}\left(\q^n,\p^n\right)\F\left(\q^{n+1/2}\right) \, ,\\
  \ell^{n+1/5} &= \ell^{n} + \xi^\Ar_{\Delta t/2, r}\left(\q^n,\p^n\right) \, ,\\
  \widetilde{\p}^{n+2/5} &= \p^{n+1/5} - (\Delta t/2)\nabla U\left(\q^{n+1/2}\right) \, ,\\
  \p^{n+2/5} &= \widetilde{\p}^{n+2/5} + \xi^\Br_{\Delta t/2, r}\left(\q^{n+1/2},\p^{n+1/5}\right)\F\left(\q^{n+1/2}\right) \, ,\\
  \ell^{n+2/5} &= \ell^{n+1/5} + \xi^\Br_{\Delta t/2, r}\left(\q^{n+1/2},\p^{n+1/5}\right) \, .
\end{align}
%where the Lagrange multiplier is defined as
%\begin{equation}
%  \xi_{r}(\q,\p) = \frac{r-\G(\q) \cdot \p}{\F(\q) \cdot \G(\q)} \, .
%\end{equation}

\noindent \textbf{Step 2}: starting from $\Delta\p^n=\mathbf{0}$, for each interacting pair within cutoff radius ($r_{ij}<r_{\mathrm{c}}$), in a successive manner,
\begin{align}
  \widetilde{\p}_i^{n+3/5} &= \p_{i}^{n+2/5}+m_{ij} \left[ \Delta v_{ij}^{\D}\left(\q^{n+1/2},\p^{n+2/5}\right) + \Delta v_{ij}^{\R}\left(\q^{n+1/2}\right) \right] \e_{ij}^{n+1/2} \, ,\\
  \widetilde{\p}_{j}^{n+3/5} &= \p_{j}^{n+2/5}-m_{ij} \left[ \Delta v_{ij}^{\D}\left(\q^{n+1/2},\p^{n+2/5}\right) + \Delta v_{ij}^{\R}\left(\q^{n+1/2}\right) \right] \e_{ij}^{n+1/2} \, ,\\ 
  \Delta\p^{n+1}_{i} &= \Delta\p^n_{i} + m_{ij} \Delta v_{ij}^{\D}\left(\q^{n+1/2},\p^{n+2/5}\right) \e_{ij}^{n+1/2} \, ,\\
  \Delta\p^{n+1}_{j} &= \Delta\p^n_{j} - m_{ij} \Delta v_{ij}^{\D}\left(\q^{n+1/2},\p^{n+2/5}\right) \e_{ij}^{n+1/2} \, , 
\end{align}
for all particles,
\begin{align}
  \p^{n+3/5} &= \widetilde{\p}^{n+3/5} + \widehat{\xi}^\Or_{\Delta t, r}\left(\q^{n+1/2},\p^{n+2/5}\right)\F\left(\q^{n+1/2}\right) \, ,\\
  \ell^{n+3/5} &= \ell^{n+2/5} - \frac{\G\left(\q^{n+1/2}\right)\cdot\Delta\p^{n+1}}{\F\left(\q^{n+1/2}\right)\cdot\G\left(\q^{n+1/2}\right)} \, .
\end{align}

\noindent \textbf{Step 3}: for all particles,
\begin{align}
  \widetilde{\p}^{n+4/5} &= \p^{n+3/5} - (\Delta t/2)\nabla U\left(\q^{n+1/2}\right) \, ,\\
  \p^{n+4/5} &= \widetilde{\p}^{n+4/5}+\xi^\Br_{\Delta t/2, r}\left(\q^{n+1/2},\p^{n+3/5}\right)\F\left(\q^{n+1/2}\right) \, ,\\
  \ell^{n+4/5} &= \ell^{n+3/5} + \xi^\Br_{\Delta t/2, r}\left(\q^{n+1/2},\p^{n+3/5}\right) \, ,\\
  \q^{n+1} &= \q^{n+1/2} + (\Delta t/2)\M^{-1}\p^{n+4/5} \, ,\\
  \p^{n+1} &= \p^{n+4/5} + \xi^\Ar_{\Delta t/2, r}\left(\q^{n+1/2},\p^{n+4/5}\right)\F\left(\q^{n+1}\right) \, ,\\
  \ell^{n+1} &= \ell^{n+4/5} + \xi^\Ar_{\Delta t/2, r}\left(\q^{n+1/2},\p^{n+4/5}\right) \, .
\end{align}
The average value of the forcing variable $\lambda$ over the corresponding time step can then be estimated as follows:
\begin{equation}
  \lambda^{n+1} = \frac{\ell^{n+1}}{\Delta t} \, ,
\end{equation}
which can be further averaged along the trajectory as
\begin{equation}
  \E^*_r[\lambda] = \frac{1}{N_\mathrm{iter}} \sum\limits_{n=1}^{N_\mathrm{iter}} \lambda^{n} \, ,
\end{equation}
where $N_\mathrm{iter}$ is the number of iterations. Note that one clear advantage of the above strategy is that we do not have to compute $\lambda$~\eqref{eq:DPD Norton lambda} along the trajectory, instead the estimation appears as natural byproducts of the integration procedure. 

\section{Numerical experiments}
\label{sec:Numerical_Experiments}

In this section, we conduct various numerical experiments to compare the performance of the DPD-NEMD system with the DPD-Norton system in computing transport coefficients in DPD, focusing on the computation of the mobility and the shear viscosity.

\subsection{Simulation details}
\label{subsec:Simulation details}

Although a soft potential (e.g., in~\cite{Shang2020}) is used in the traditional formulation of DPD, alternative potentials have also been widely used in DPD. In this article, we consider the following Lennard-Jones total potential energy, which was also used in~\cite{Blassel2023}: 
\begin{equation}
  %U(\mathbf{q}) = \sum\limits_{i=1}^{N-1}\sum\limits_{j=i+1}^{N}1/2a_{ij}r_{ij}\left(1-\frac{r_{ij}}{r_{\mathrm{c}}}\right)^{2}\mathbf{1}_{\{r_{ij}<r_{\mathrm{c}}\}},
  U(\mathbf{q}) = \sum\limits_{i=1}^{N-1}\sum\limits_{j=i+1}^{N} \varphi\left(r_{ij}\right) \, ,
\end{equation}
where $\varphi\left(r_{ij}\right)$ represents the pair potential energy, obtained by truncating the range of $u$ to ensure $\varphi$ is $C^{1}$ (i.e., continuously differentiable),
\begin{equation}
\varphi\left(r_{ij}\right)=
  \begin{cases} u\left(r_{ij}\right)-u\left(r_{\mathrm{c}}\right)-u'\left(r_{\mathrm{c}}\right)\left(r_{ij}-r_{\mathrm{c}}\right) \, , &r_{ij}<r_{\mathrm{c}} \, ,\\
    \quad \quad \quad \quad \quad 0 \, , & r_{ij}\geq r_{\mathrm{c}} \, ,
  \end{cases}
\end{equation}
where $u$ is given by
\begin{equation}
u(r)=4\varepsilon\left[\left(\frac{r_{0}}{r}\right)^{12}-\left(\frac{r_{0}}{r}\right)^6\right] \, ,
\end{equation}
%\begin{equation}
%    %U(\mathbf{q}) = \sum\limits_{i=1}^{N-1}\sum\limits_{j=i+1}^{N}1/2a_{ij}r_{ij}\left(1-\frac{r_{ij}}{r_{\mathrm{c}}}\right)^{2}\mathbf{1}_{\{r_{ij}<r_{\mathrm{c}}\}},
%    \varphi\left(r_{ij}\right) = \sum\limits_{i=1}^{N-1}\sum\limits_{j=i+1}^{N}4\varepsilon\left[\left(\frac{r_{\mathrm{c}}}{r_{ij}}\right)^{12}-\left(\frac{r_{\mathrm{c}}}{r_{ij}}\right)^6\right]\mathbf{1}_{\{r_{ij}<r_{\mathrm{c}}\}}\, ,
%\end{equation}
where $\varepsilon$ and $r_{0}$ are two constants that set the energy and length scales
of the beads, respectively, in reduced Lennard-Jones units. It is worth mentioning that we have also tested the soft potential (e.g., in~\cite{Shang2020}) and the results are very similar to the Lennard-Jones potential, and only the results of the Lennard-Jones potential are presented.

In our numerical experiments, a system of $N=500$ identical particles was simulated in a cubic box (i.e., $d=3$) with periodic boundary conditions, unless otherwise stated. The following parameter set was used: $\varepsilon=r_{0}=m_{i}=1$ (for defining reduced units), and $k_{\mathrm{B}}=T=1$. Particle density $\rho_{\rm d}=0.85$ was used with cutoff radius $r_{\mathrm{c}}=2.5$. The particles are initiated on a cubic lattice with equal spacing, while the initial momenta were independent and identically distributed normal random variables with mean zero and variance $k_{\mathrm{B}}T$. We included the case of $\gamma=4.5$ for a thorough investigation, and we also compared the performance with a friction coefficient of $\gamma=40.5$.
%among various values of the friction coefficients, especially $\gamma=1.0$, $\gamma=4.5$ and $\gamma=40.5$ that were widely used in algorithms tests in DPD. 
Cell lists~\cite{Verlet1967} were used in each numerical simulation.

Unless otherwise stated, the system was simulated for 1000 reduced time units, with a stepsize of $\Delta t=0.01$, but only the last 80\,\% of the data were collected to calculate the physical quantities in order to make sure the system was well equilibrated---we have numerically verified that this was indeed the case. Ten different runs were averaged to reduce the sampling errors. All the numerical simulations were performed on the University of Birmingham supercomputer \href{https://www.birmingham.ac.uk/research/arc/bear}{BlueBEAR} by using the Julia package \texttt{Molly}~\cite{Greener2024}.

\subsection{Transport coefficients}
\label{subsec:Transport coefficients}

In this subsection, we introduce two transport coefficients (i.e., the mobility and the shear viscosity) that will be computed and compared in our numerical results in Section~\eqref{subsec:Numerical results}.

\subsubsection{Mobility}
\label{subsubsec:Transport Mobility}

Proportional to the diffusion coefficient, mobility is an important transport coefficient in physics, and has been widely studied for the Langevin dynamics~\cite{Evans2008,Wang2013a,Leimkuhler2013c,Pavliotis2022,Spacek2023}. The response $R$ associated with the mobility is given by
\begin{equation}\label{eq:mobility response}
  R(\q,\p) = \G(\q) \cdot \p = \F \cdot \M^{-1}\p \, ,
\end{equation}
where the external force $\F(\q) = \F = \left(\F_1,\F_2,\cdots,\F_N\right)^\TT \in \mathbb{R}^{3N}$ is a constant vector field and we further assume that it is normalised such that $|\F|=1$. We consider two types of drifts as follows~\cite{Evans2008}:
\begin{itemize}
  \item \textbf{Two drifts}: this approach corresponds to a perturbation where the force acts on one particle in one direction, and on another particle in the opposite direction, which can be assumed (by indistinguishability of the particles) to be the $x$ component of the first two particles
      \begin{equation}
        \F_1 = \frac{1}{\sqrt{2}} \left(1, 0, 0\right)^\TT \, , \quad \F_2 = \frac{1}{\sqrt{2}} \left(-1, 0, 0\right)^\TT \, , \quad \F_i = \mathbf{0} \, , \quad i=3,4,\ldots,N \, ,
      \end{equation}
  \item \textbf{Colour drifts}: this approach corresponds to a perturbation where the force acts on one half of the particles in one direction, and on another half of the particles in the opposite direction, which we again choose by convention to be the $x$ direction
      \begin{equation}
        \F_i = \frac{1}{\sqrt{N}} \left( (-1)^{i}, 0, 0 \right)^\TT, \quad i=1,2,\ldots,N \, ,
      \end{equation}
      where we assume $N$ is an even number to ensure the momentum conservation.
\end{itemize}
It is worth mentioning that the controllability condition $\G(\q) \cdot \F(\q) = \F \cdot \M^{-1}\F \neq 0$ is satisfied in both cases described above.

Based on the Einstein relation~\cite{Rodenhausen1989}, the mobility $\alpha$ is related to the diffusion coefficient $D$ as follows:
\begin{equation}\label{eq:Einstein}
  \alpha = \beta D \, .
\end{equation}
In addition to the two popular approaches (i.e., Green--Kubo and NEMD) to measure the transport coefficients, another common method to approximate the diffusion coefficient is the so-called ``mean squared displacement (MSD)''. The MSD at time $t$ is defined as an ensemble average
\begin{equation}
  \mathrm{MSD}(t) = \left\langle \left|\q(t)-\q(0)\right|^2 \right\rangle = \lim_{N\rightarrow\infty} \frac{1}{N} \sum\limits_{i=1}^{N} \left|\q_{i}(t)-\q_{i}(0)\right|^2 \, .
\end{equation}
The diffusion coefficient can then be approximated as follows: 
\begin{equation}\label{eq:msd}
  D = \frac{1}{2d} \lim_{t\rightarrow\infty} \frac{\mathrm{MSD}(t)}{t} \, .
\end{equation}
The MSD approach has been widely adopted in DPD~\cite{Vattulainen2002,Chaudhri2010,Panoukidou2021}, and in what follows we will use the approximation obtained from the MSD approach via~\eqref{eq:Einstein} as the reference value for the comparison of the mobility approximations.

\subsubsection{Shear viscosity}
\label{subsubsec:Transport Shear viscosity}

Shear viscosity is another important transport coefficient in physics. A common approach to computing the shear viscosity in DPD is to apply Lees--Edwards boundary conditions in nonequilibrium simulations to generate a steady shear flow~\cite{Pagonabarraga1998,Shang2020}, whereas an alternative method in equilibrium simulations is to use the Green--Kubo integration of the stress autocorrelation function~\cite{Chaudhri2010,Panoukidou2021}. Based on the linear response theory, a rigorous framework for computing the shear viscosity in two-dimensional Langevin dynamics was developed in~\cite{Joubaud2012}, and the corresponding numerical investigation in three dimensions, including an extension to Norton dynamics, was carried out in~\cite{Blassel2023}. In this article, we present a detailed mathematical derivation demonstrating how the shear viscosity of a three-dimensional DPD system can be obtained from the average longitudinal
velocity.

Denote the lengths of the DPD system in each spatial direction as $L_x$, $L_y$, and $L_z$, respectively, we consider an external force acting in the longitudinal $x$-direction, with its magnitude determined by the transverse configuration in the $y$-direction. Specifically, three types of external forces are examined in this article, as defined in~\cite{Joubaud2012,Blassel2023}:
\begin{itemize}
  \item sinusoidal force: $F(Y)=\sin\left(\frac{2\pi Y}{L_y}\right)$,
  \item piecewise linear force: $F(Y)=
      \begin{cases}
        \frac{4}{L_y}\left(Y-\frac{L_y}{4}\right)\, ,&0\leq Y \leq\frac{L_y}{2}\, ,\\
        \frac{4}{L_y}\left(\frac{3L_y}{4}-Y\right)\, ,&\frac{L_y}{2}< Y \leq L_y\, ,\\
      \end{cases}$
  \item piecewise constant force: $F(Y)=
      \begin{cases}
        1\, ,&0\leq Y \leq\frac{L_y}{2}\, ,\\
        -1\, ,&\frac{L_y}{2}< Y \leq L_y\, .\\
      \end{cases}$
\end{itemize}
The average longitudinal velocity $U_x^\varepsilon(Y,\q,\p)$, the $xy$ component of the stress tensor $\Sigma_{xy}^\varepsilon(Y,\q,\p)$ are defined, respectively as follows:
\begin{subequations}
\begin{align}
  U_x^\varepsilon(Y,\q,\p) &= \frac{L_y}{Nm} \sum\limits_{i=1}^N p^{x}_{i} \chi_\varepsilon\left(q^{y}_{i}-Y\right) \, ,\\
  \Sigma_{xy}^\varepsilon(Y,\q,\p) &= \frac{1}{L_x L_z} \left( \sum\limits_{i=1}^N\frac{p^{x}_{i}p^{y}_{i}}{m}\chi_\varepsilon\left(q^{y}_{i}-Y\right) + \sum\limits_{1\leq i<j\leq N}F_{ij}^x\left(r_{ij}\right)\int_{q^{y}_{j}}^{q^{y}_{i}}\chi_\varepsilon(s-Y) \, \dd s \right) \, , \label{eq:stress}
\end{align}
\end{subequations}
where $\chi_\varepsilon$ is a smooth approximation of the Dirac delta function on $\left[0,L_y\right]$ with $0 <\varepsilon\leq 1$ (see more details in~\cite{Espanol1995a, Joubaud2012}), and the intermolecular force $F_{ij}^x\left(r_{ij}\right)$, which includes both conservative and dissipative contributions in the case of DPD~\cite{Espanol1995a}, is given by
\begin{equation}
    F_{ij}^x\left(r_{ij}\right)=-U'\left(r_{ij}\right)e_{ij}^x-\gamma\omega^{\D}\left(r_{ij}\right)\left(\mathbf{e}_{ij}\cdot\mathbf{v}_{ij}\right)e_{ij}^x\, ,
\end{equation}
where $e_{ij}^x$ is the $x$ component of $\e_{ij}$.
%$\e_{ij}=(e_{ij}^x,e_{ij}^y,e_{ij}^z)$.

In analogy with the case of Langevin dynamics proposed in~\cite{Joubaud2012}, we begin by analysing the linear response of the DPD system with small perturbations. To derive a closed-form expression for the shear viscosity, we consider the average longitudinal velocity and the off-diagonal term of the stress tensor in the limits of vanishing forcing and spatial averaging in the context of DPD, and present the following proposition.

\begin{proposition}
Suppose that the limits
\begin{subequations}
\begin{align}
  u_x(Y) & =\lim\limits_{\varepsilon\rightarrow0} \lim\limits_{\eta\rightarrow0} \frac{ \left\langle U_x^\varepsilon(Y,\q,\p)\right\rangle_{\eta} }{ \eta } \, ,\\
  \sigma_{xy}(Y) &= \lim\limits_{\varepsilon\rightarrow0} \lim\limits_{\eta\rightarrow0} \frac{ \left\langle\Sigma_{xy}^\varepsilon(Y,\q,\p)\right\rangle_{\eta} }{ \eta } \, ,
\end{align}
\end{subequations} 
exist and are smooth with respect to $Y\in\left[0,L_y\right]$, where $\langle \cdot \rangle_{\eta}$ denotes the average associated with the measure of the dynamics~\eqref{eq:DPD NEMD}, then
    \begin{equation}
        \frac{\partial\sigma_{xy}(Y)}{\partial Y}=\rho F(Y)\, ,
        \label{eq:prop1}
    \end{equation}
    where $\rho=\frac{N}{L_xL_yL_z}$ is the density and $F(Y)$ is the external force.
    \label{prop1}
\end{proposition}

The proof of Proposition~\ref{prop1} is provided in~\ref{appendix:prop1}. In this article, we consider a bulk homogeneous system and assume that the external force is sufficiently small to remain within the linear response regime, in which the fluid exhibits the Newtonian behaviour with constant shear viscosity. Hence, by the definition of shear viscosity, denoted by $\nu$, we obtain the following relation:
\begin{equation}
    \sigma_{xy}(Y)=-\nu\frac{\dd u_x(Y)}{\dd Y}\, ,
\end{equation}
from which~\eqref{eq:prop1} can be rewritten as
\begin{equation}
    \nu u_x^{\prime\prime}(Y)=-\rho F(Y)\, .\label{eq:shear}
\end{equation}
Finally, the shear viscosity can be computed by applying a Fourier transform to~\eqref{eq:shear}, yielding
\begin{equation}
    \nu=\frac{\rho F_1}{U_1}\left(\frac{L_y}{2\pi}\right)^2\, ,\label{eq:shear viscosity computation}
\end{equation}
where $F_1$ and $U_1$ denote the first Fourier coefficient of $F$ and $u_x$, respectively, as follows:
\begin{equation}
  U_1 = \frac{1}{L_y}\int_0^{L_y}u_x(Y)e^{\frac{2i\pi Y}{L_y}} \, \dd Y \, , \quad F_1 = \frac{1}{L_y} \int_0^{L_y} F(Y)e^{\frac{2i\pi Y}{L_y}} \, \dd Y \, .
\end{equation}
In estimating the shear viscosity, $F_1$ is analytically known, and $U_1$ can be computed by the corresponding response $R$ as follows:
\begin{equation}
  R(\q,\p) = \F\left(\q_y\right) \cdot \M^{-1}\p = \frac{1}{mN}\sum\limits^{N}_{j=1}p^{x}_{j}\exp\left(\frac{2i\pi q^{y}_{j}}{L_y}\right)\, .
\end{equation}

\subsection{Asymptotic variances}
\label{subsec:Asymptotic variances} 

The challenge associated with NEMD methods is that the estimator~\eqref{eq:NEMD transport estimator} suffers from large statistical errors when $|\eta|$ is small. More precisely, we can show that the asymptotic variance of the estimator~\eqref{eq:NEMD transport estimator} scales as $\eta^{-2}$ as $\eta$ approaches zero~\cite{Lelievre2016,Blassel2023}. Although there exist techniques to extend the linear regime in NEMD methods~\cite{Spacek2023}, in many cases the linear regimes are limited, resulting in large asymptotic variances. Furthermore, we have
\begin{equation}\label{eq:NEMD var}
  \sigma^2_{\alpha,\eta} = \frac{2}{\eta^2} \Var_\eta(R) \Theta_\eta(R) \, ,
\end{equation}
where $\Var_\eta$ denotes the variance with respect to $\E_\eta$ and the correlation time $\Theta_\eta(R)$ is given by
\begin{equation}\label{eq:NEMD auto}
  \Theta_\eta(R) = \int_0^\infty \frac{\E_\eta\left[\Pi_{\eta}R\left(\q_t,\p_t\right) \Pi_{\eta}R\left(\q_0,\p_0\right) \right]}{\E_\eta\left[(\Pi_{\eta}R)^2\right]} \, \dd t \, ,
\end{equation}
where $\Pi_{\eta}$ is the centering operator given by $\Pi_{\eta}\varphi = \varphi - \E_\eta[\varphi]$.

Similar to the analyses in NEMD methods, we can show that the asymptotic variance of the estimator~\eqref{eq:Norton transport estimator} scales as $r^{-2}$ as $r$ approaches zero~\cite{Blassel2023}. More precisely, we have
\begin{equation}\label{eq:Norton var}
  \sigma^2_{\alpha^*, r} = \frac{2r^2}{(\E^*_r[\lambda])^4} \Var_r^*(\lambda) \Theta_r^*(\lambda) = \frac{2(\alpha^*)^4}{r^2} \Var_r^*(\lambda) \Theta_r^*(\lambda) + O \left( \frac{1}{r} \right) \, ,
\end{equation}
where $\Var_r^*$ denotes the variance with respect to $\E^*_r$ and the correlation time $\Theta_r^*(\lambda)$ is defined similarly to the NEMD case~\eqref{eq:NEMD auto} by
\begin{equation}\label{eq:Norton auto}
  \Theta^*_r(\lambda) = \int_0^\infty \frac{\E^*_r\left[ \Pi^*_r \lambda\left(\q_t,\p_t\right) \Pi^*_r \lambda\left(\q_0,\p_0\right) \right]}{\E^*_r\left[ (\Pi^*_r \lambda)^2 \right]} \, \dd t \, ,
\end{equation}
where $\Pi^*_r$ is the centering operator with respect to $\E^*_r$.

\subsection{Numerical results}
\label{subsec:Numerical results} 

In this subsection, we present our numerical results on the mobility and the shear viscosity.

\subsubsection{Mobility}
\label{subsubsec:Mobility}

\begin{figure}[tbp]
  \centering
  \includegraphics[scale=0.38]{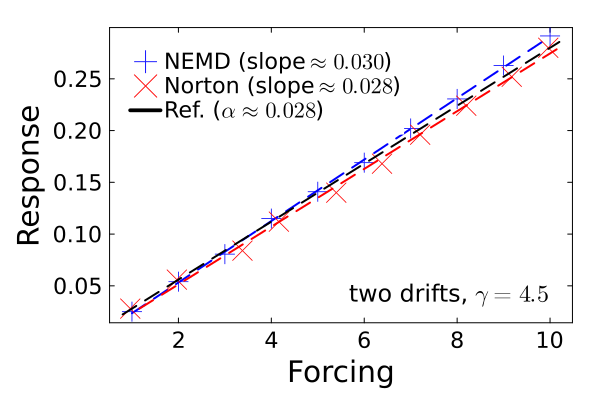}
  \includegraphics[scale=0.38]{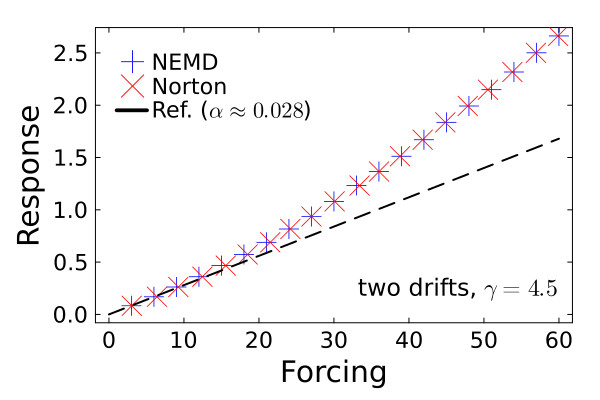}
  \includegraphics[scale=0.38]{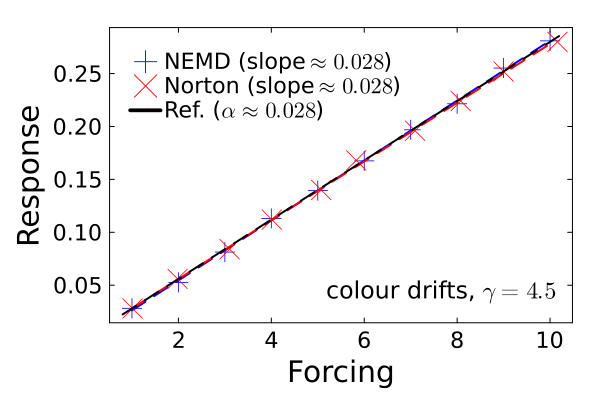}
  \includegraphics[scale=0.38]{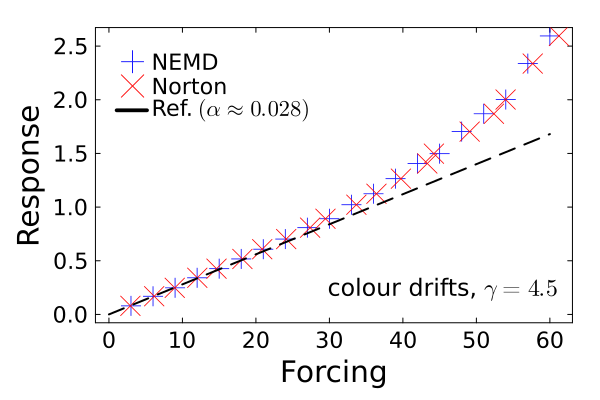}
  \caption{\small Response profiles of the mobility in the linear (left) and nonlinear (right) regimes for the DPD-NEMD~\eqref{eq:DPD NEMD} and DPD-Norton~\eqref{eq:DPD Norton} dynamics with a friction coefficient of $\gamma=4.5$ and two types of external forces (top: two drifts; bottom: colour drifts). The scattered markers represent the experimental results, and the slope of the dashed linear regression lines corresponds to the estimated mobility (estimations are shown in parentheses). The approximation obtained from the MSD approach is viewed as the reference value.}
  \label{fig:mobility gamma=4.5}
\end{figure}

In Figure~\ref{fig:mobility gamma=4.5}, we plot the response as a function of the forcing magnitude for both the DPD-NEMD and DPD-Norton dynamics with a friction coefficient of $\gamma=4.5$. Therefore, the fixed quantity is plotted in the horizontal axis for the DPD-NEMD dynamics, and in the vertical axis for the DPD-Norton dynamics, for the cases of two drifts (top row) and colour drifts (bottom row), respectively. In the linear regime, shown on the left panels, the slope of the dashed linear regression lines corresponds to the estimated mobility, while the approximation obtained from the MSD approach is also plotted as the reference value. We observe that, in the two drifts case, the results from DPD-Norton exhibit better agreement with the reference value, when the estimations from DPD-NEMD show a mild deviation. In contrast, in the colour drifts case, the results from both methods are almost indistinguishable, both in excellent agreement with the reference value. These results suggest that, with little computational overhead, DPD-Norton is more accurate than DPD-NEMD in the two drifts case, whereas both methods produce very similar excellent estimates in the colour drift case. The right panels compare the responses with a much wider range of the values of the forcing magnitude until $\eta=60$ for both cases. It is shown that in the two drifts case the response clearly starts to display nonlinear behaviour with a forcing magnitude of $\eta>15$, whereas the response still appears to be linear up until a forcing magnitude of $\eta=25$ in the colour drifts case. This again demonstrates the superiority of the colour drifts in maintaining a linear regime for a much wider range of the values of the forcing magnitude.

%\begin{figure}[tp]
%  \centering
%  \includegraphics[scale=0.38]{mobility/fig7colour_eta5run10sim1000gamma4_5dt01V1.png}
%  \includegraphics[scale=0.38]{mobility/fig7colour_r_14run10sim1000gamma4_5dt01V1.png}
%  \caption{\small Boxplot of the mobility as a function of the system size with the colour drifts and a friction coefficient of $\gamma = 4.5$, using $\eta=5.0$ for the DPD-NEMD~\eqref{eq:DPD NEMD} (left) and $r=0.14$ for the DPD-Norton dynamics~\eqref{eq:DPD Norton} (right).}
%\end{figure}

\begin{figure}[tbp]
  \centering
  \includegraphics[scale=0.38]{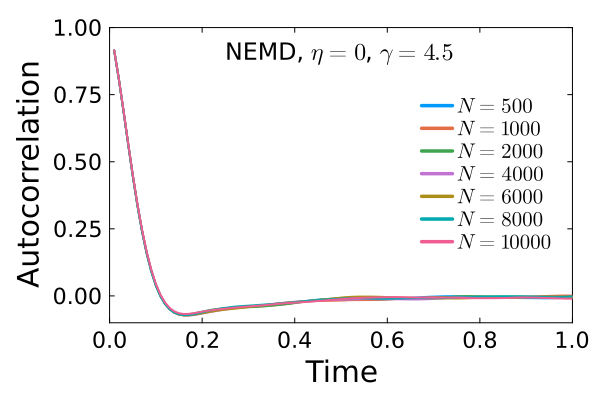}
  \includegraphics[scale=0.38]{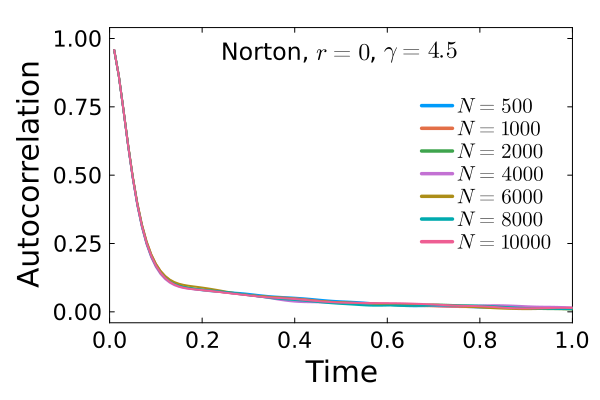}
  \caption{\small Autocorrelation functions in equilibrium for the DPD-NEMD~\eqref{eq:NEMD auto} (left) and DPD-Norton~\eqref{eq:Norton auto} (right) dynamics, corresponding to the response observable $R$~\eqref{eq:mobility response} and the forcing observable $\lambda$~\eqref{eq:DPD Norton lambda}, respectively. Simulations were performed with a friction coefficient of $\gamma = 4.5$ and a wide range of system sizes $N$.}
  \label{fig:mobility auto}
\end{figure}

Figure~\ref{fig:mobility auto} compares the autocorrelation functions of the response observable $R$ and the forcing observable $\lambda$ in equilibrium, corresponding to $\eta = 0$ for the DPD-NEMD dynamics~\eqref{eq:NEMD auto} and $r = 0$ for the DPD-Norton dynamics~\eqref{eq:Norton auto}, with a friction coefficient of $\gamma = 4.5$ and a wide range of system sizes $N$. It is observed that the autocorrelation functions decay very rapidly and the correlation times appear to be very small. We would like to emphasise that the aim of the figure is to show that each observable is independent of $N$. The autocorrelation functions with a wide range of system sizes are almost indistinguishable, indicating that the system size has little effect on the computation of the mobility in DPD---we have numerically verified that this is indeed the case.

\begin{figure}[tp]
  \centering
  \includegraphics[scale=0.38]{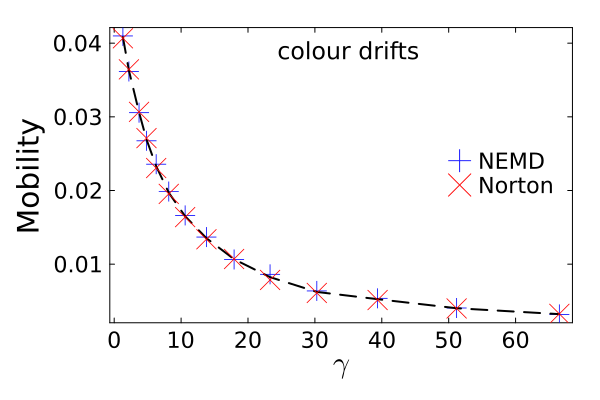}
  \includegraphics[scale=0.38]{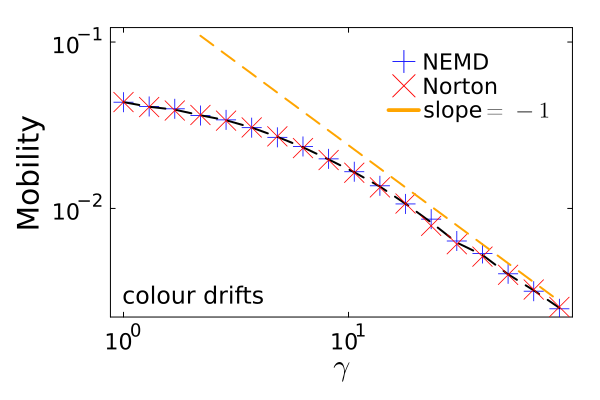}
  \caption{\small Estimated mobility as a function of the friction coefficient $\gamma$ (left: linear scale; right: log--log scale) in the colour drifts case, by using the DPD-NEMD~\eqref{eq:DPD NEMD} and DPD-Norton~\eqref{eq:DPD Norton} dynamics.}
  \label{fig:mobility gamma}
\end{figure}

The estimated mobility with a wide range of friction coefficients $\gamma$, from 1 to approximately 86, for both the DPD-NEMD and DPD-Norton dynamics in the colour drifts case is plotted in Figure~\ref{fig:mobility gamma}. The results show a clear monotonic decrease in mobility as $\gamma$ increases, where the results from both dynamics are almost indistinguishable. The log--log plot on the right panel illustrates that, as $\gamma$ increases to large values, the contribution from the dissipative force becomes dominant, which is consistent with the results from the ideal fluids as shown in~\cite{Groot1997}, where the diffusion coefficient scales as $D \propto \gamma^{-1}$, in accordance with its relation to the mobility via the Einstein relation~\eqref{eq:Einstein}. However, for small values of $\gamma$, the contribution from the conservative force is not negligible. In addition, by switching off the conservative force, we have verified our results in the case of ideal fluids---they are in good agreement with those obtained in~\cite{Groot1997}. 
%Moreover, we have numerically confirmed that the mobility is independent of the system size, as shown in Appendix~\ref{appendix:fig}.

%This is consistent with the inverse relation between the diffusion and the friction coefficient in DPD~\cite{Groot1997}, however, the values of mobility observed in our results, which equal to the values of the diffusion due to $\beta=1$, do not follow an exact inverse proportionality with respect to $\gamma$. 

\begin{figure}[tp]
  \centering
  \includegraphics[scale=0.38]{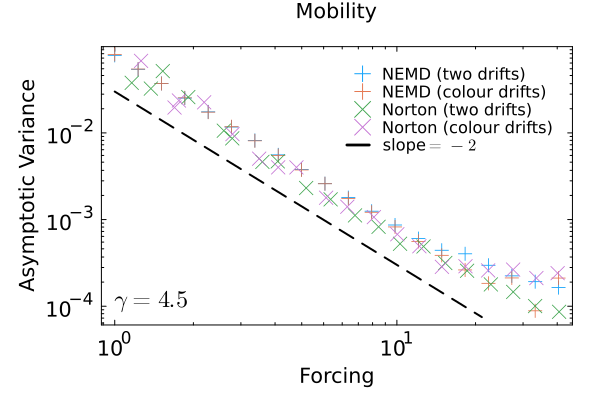}
  \includegraphics[scale=0.38]{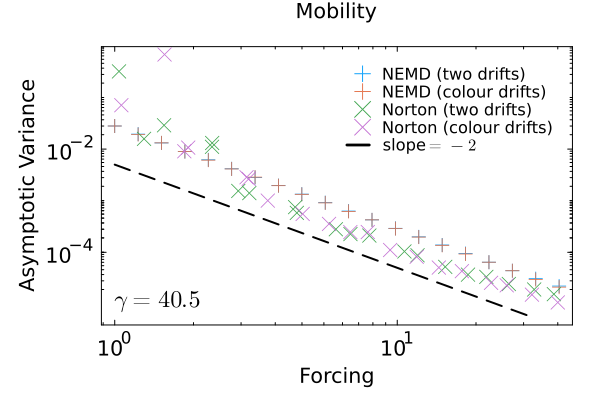}
  \caption{\small Double logarithmic plot of the asymptotic variance for the estimators of the mobility against the forcing magnitude for the DPD-NEMD~\eqref{eq:DPD NEMD} and DPD-Norton~\eqref{eq:DPD Norton} dynamics by using the two drifts and colour drifts, with friction coefficients of 
  %$\gamma=1$ (top left), 
  $\gamma=4.5$ (left) and $\gamma=40.5$ (right). The expected scaling line for small values of the forcing or response is plotted in dashed lines in all cases.}
  \label{fig:mobility var}
\end{figure}

With the help of the block averaging method~\cite{Flyvbjerg1989}, Figure~\ref{fig:mobility var} compares the asymptotic variance, a key metric to measure the associated computational costs, for the estimators of the mobility against the forcing magnitude in a log--log plot, with friction coefficients of $\gamma=4.5$ and $\gamma=40.5$. It is worth mentioning that the horizontal axes represent the forcing $\eta$ in the DPD-NEMD dynamics and the average value of the forcing observable $\lambda$ in the DPD-Norton dynamics, respectively. Note that the asymptotic variances indeed scale as $\eta^{-2}$ for the DPD-NEMD dynamics, and $r^{-2}$ for the DPD-Norton dynamics in the linear regime. It can be seen from the left panel that the asymptotic variances appear to depart from the expected order of $-2$ when the forcing magnitudes are relatively large for both dynamics; this may not be that surprising as we can see from Figure~\ref{fig:mobility gamma=4.5} that when the forcing magnitudes are relatively large the responses may not be in the linear regime.
%, at least for small values of $|\eta|$ and $|r|$. 
While the results obtained from the two drifts are very similar to those obtained from the colour drifts in most cases particularly in the linear regime, DPD-Norton appears to have an improved asymptotic variance particularly with a large value of the friction coefficient, indicating an improvement in the statistical efficiency of DPD-Norton over DPD-NEMD. With 
%a friction coefficient of $\gamma = 1$, the asymptotic variances of both dynamics are very similar to each other when the forcing magnitude is not over 10, while with 
a friction coefficient of $\gamma = 4.5$, the asymptotic variance associated with the DPD-Norton dynamics appears to be slightly better than that of DPD-NEMD, particularly when the forcing magnitude is relatively small. It is worth noting that the DPD-Norton dynamics clearly outperforms the DPD-NEMD dynamics with a friction coefficient of $\gamma = 40.5$. Moreover, in the high friction limit, the asymptotic variances retain the expected order of $-2$ even when the forcing magnitudes are relatively large for both dynamics, although the asymptotic variance appears to be unstable when the forcing magnitudes are very small (we have checked the intermediate cases of $\gamma = 18$ and $\gamma = 32$, and observed that the deviation from the expected order of $-2$, when the forcing magnitudes are very small, becomes stronger as we increase $\gamma$, indicating that larger forcing magnitudes may be needed in the higher friction limits).
% In particular, in the case of $\gamma=4.5$ on the right panel, the asymptotic variances obtained from the DPD-Norton dynamics are almost one magnitude smaller than those obtained from the DPD-NEMD dynamics when the forcing magnitude is small. 

\begin{figure}[tbp]
  \centering
  \includegraphics[scale=0.38]{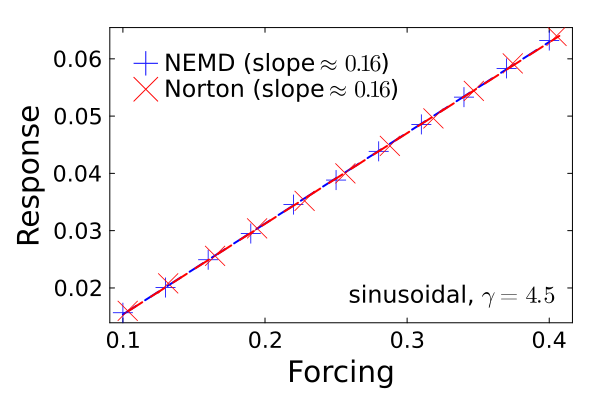}
  \includegraphics[scale=0.38]{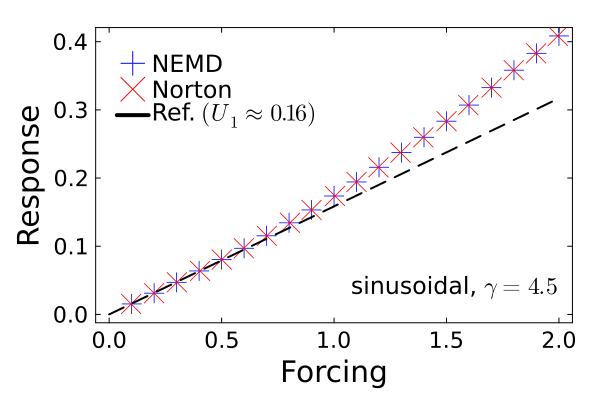}
  \includegraphics[scale=0.38]{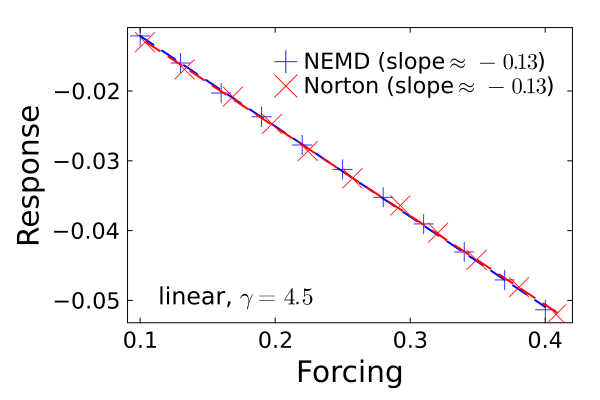}
  \includegraphics[scale=0.38]{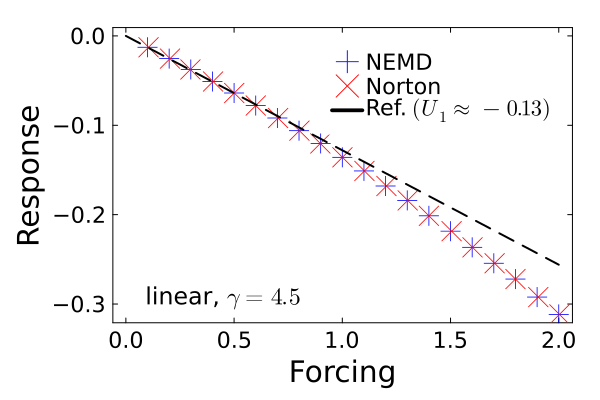}
  \includegraphics[scale=0.38]{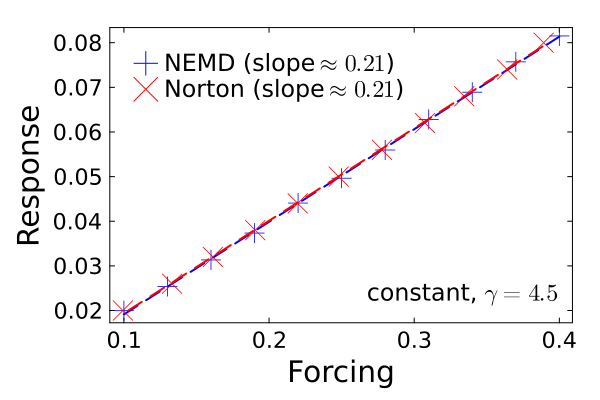}
  \includegraphics[scale=0.38]{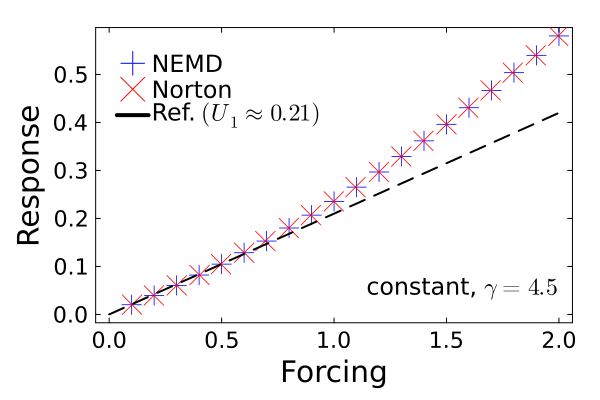}
  \caption{\small Response profiles of the Fourier coefficient $U_1$ associated with the shear viscosity in the linear (left) and nonlinear (right) regimes for the DPD-NEMD~\eqref{eq:DPD NEMD} and DPD-Norton~\eqref{eq:DPD Norton} dynamics with a friction coefficient of $\gamma=4.5$ and three types of external forces (top: sinusoidal force; middle: piecewise linear force; bottom: piecewise constant force). The format of the plots is the same as in Figure~\ref{fig:mobility gamma=4.5}.}
  \label{fig:shear viscosity gamma=4.5}
\end{figure}

\subsubsection{Shear viscosity}
\label{subsubsec:Shear viscosity}

\begin{figure}[tbp]
  \centering
  \includegraphics[scale=0.38]{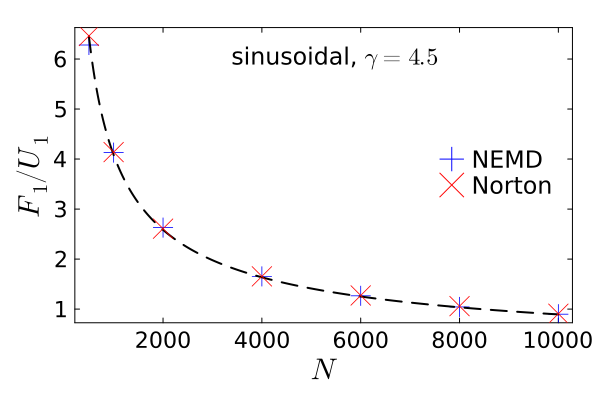}
  \includegraphics[scale=0.38]{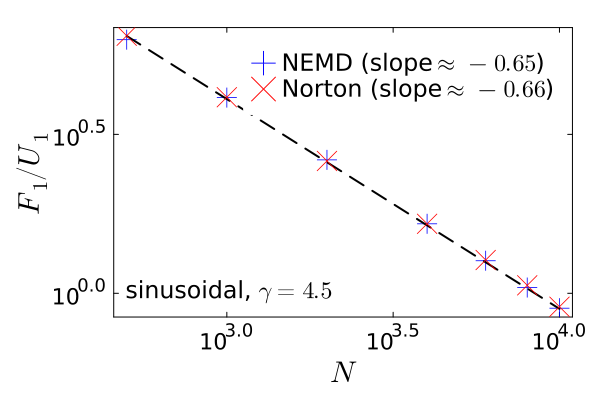}
  \caption{\small Estimator $F_1 / U_1$ as a function of the system size $N$ (left: linear scale; right: log--log scale) with the sinusoidal force and a friction coefficient of $\gamma = 4.5$, using $\eta=0.1$ for the DPD-NEMD~\eqref{eq:DPD NEMD} and $r=0.016$ for the DPD-Norton dynamics~\eqref{eq:DPD Norton}.}
  \label{fig:shear viscosity F1/U1}
\end{figure}

\begin{figure}[tbp]
  \centering
  \includegraphics[scale=0.38]{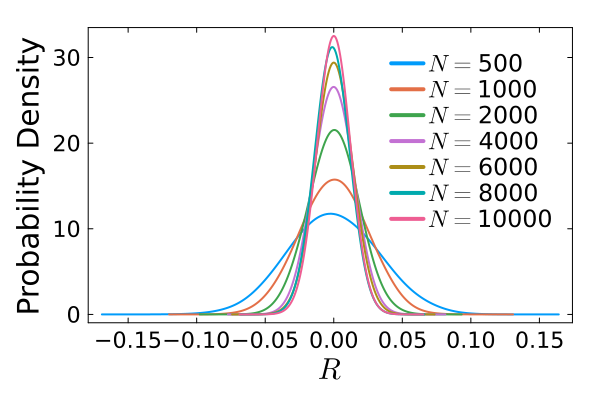}
  \includegraphics[scale=0.38]{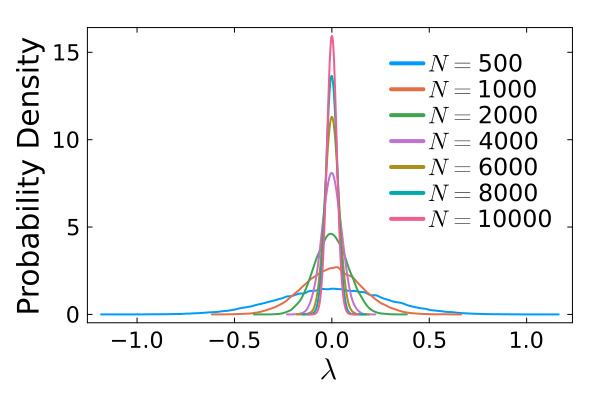}
  \caption{\small Histograms of the response observable $R$ (left) and the forcing observable $\lambda$ (right) in equilibrium in the DPD-NEMD~\eqref{eq:DPD NEMD} and DPD-Norton~\eqref{eq:DPD Norton} dynamics, respectively, with the sinusoidal force and $\gamma = 4.5$ and a wide range of system sizes $N$.}
  \label{fig:histogram}
\end{figure}

\begin{figure}[tbp]
  \centering
  \includegraphics[scale=0.38]{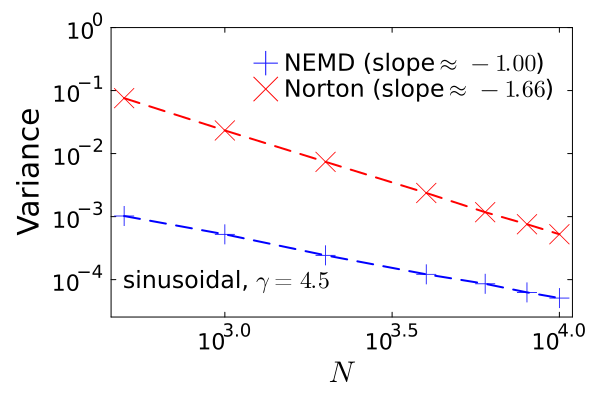}
  \caption{\small Variance of the response observable $R$ and the forcing observable $\lambda$ as a function of the system size $N$ in a log--log scale, obtained from the histograms in Figure~\ref{fig:histogram}.}
  \label{fig:variance}
\end{figure}

We now proceed to investigate the computation of the shear viscosity with various types of external forces as defined in~\cite{Joubaud2012,Blassel2023} and compare the performance of estimations obtained from both the DPD-NEMD and DPD-Norton dynamics.

Figure~\ref{fig:shear viscosity gamma=4.5} displays the numerical estimations of the Fourier responses $U_1$ of the forcing magnitude for both the DPD-NEMD and DPD-Norton dynamics with a friction coefficient of $\gamma=4.5$, and the sinusoidal, piecewise linear, and piecewise constant forces, respectively. The values of shear viscosity can be computed from~\eqref{eq:shear viscosity computation}, where $U_1$ is estimated from the slope of the response with respect to the forcing magnitude, and the Fourier coefficient $F_1$ is analytically known that $F_1=i/2$ for the sinusoidal force, $F_1=-4/\pi^2$ for the piecewise linear force, and $F_1=2i/\pi$ for the piecewise constant force. Note that in the numerical implementation, we compute the imaginary part of the Fourier coefficient for the sinusoidal and piecewise constant forces, while the real part is considered for the piecewise linear force as the corresponding imaginary component vanishes. For all three types of forces, the results obtained from the DPD-NEMD and DPD-Norton dynamics are almost indistinguishable, leading to consistent shear viscosity estimates in the linear regime and almost identical behaviour in the nonlinear regime. 

The estimators $F_1 / U_1$ with the sinusoidal force are plotted in Figure~\ref{fig:shear viscosity F1/U1}, as a function of the system size $N$, for both the DPD-NEMD and DPD-Norton dynamics. As $N$ increases, the estimators exhibit approximately linear behaviour on a log--log scale. This confirms a power-law convergence behaviour for the estimator, with the scaling observed to be asymptotically $N^{-2/3}$, consistent with the expression of the shear viscosity~\eqref{eq:shear viscosity computation}. 

%The results from the two dynamics are nearly indistinguishable across all tested system sizes, further validating the robustness of the proposed estimator with respect to discretization.

To further investigate the equilibrium fluctuation behaviour in the thermodynamic limit, Figure~\ref{fig:histogram} shows the histograms of the response observable $R$ for the DPD-NEMD dynamics and the forcing observable $\lambda$ for the DPD-Norton dynamics with the sinusoidal force, and various values of $N$ in equilibrium (i.e., $\eta = 0$ and $r = 0$). As $N$ increases, both histograms become increasingly concentrated around zero, indicating reduced variance of the observables in the thermodynamic limit. This behaviour is further confirmed quantitatively in Figure~\ref{fig:variance}, where the variances of $R$ and $\lambda$ are plotted against $N$ in a log--log scale. Moreover, the asymptotic variance of $R$ scales as $N^{-1}$, while the asymptotic variance of $\lambda$ decays significantly faster, as $N^{-5/3}$.

\begin{figure}[tbp]
  \centering
  \includegraphics[scale=0.38]{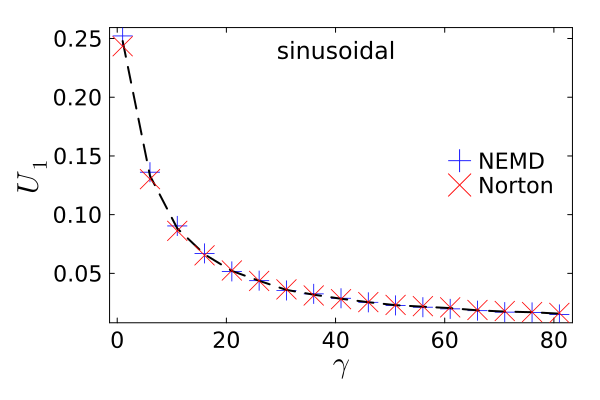}
  \includegraphics[scale=0.38]{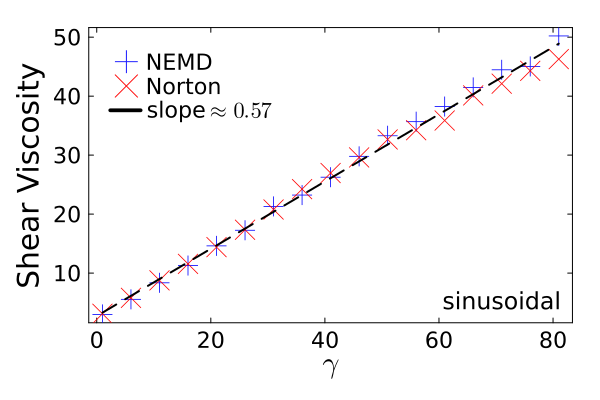}
  \caption{\small Left: Estimated Fourier response $U_1$ as a function of the friction coefficient $\gamma$ with the sinusoidal force for the DPD-NEMD~\eqref{eq:DPD NEMD} and DPD-Norton~\eqref{eq:DPD Norton} dynamics. Right: Corresponding shear viscosity computed via~\eqref{eq:shear viscosity computation}.}
  \label{fig:shear viscosity gamma}
\end{figure}

In Figure~\ref{fig:shear viscosity gamma}, we plot the estimated Fourier response $U_1$ (left) and the corresponding shear viscosity $\nu$ (right) as a function of the friction coefficient $\gamma$ with the sinusoidal force, for both the DPD-NEMD and DPD-Norton dynamics. The results from both dynamics are almost identical with a range of $\gamma$, confirming their consistency in capturing the friction-dependent transport behaviour. The response $U_1$ decreases monotonically as $\gamma$ increases, resulting in a corresponding increase in the computed shear viscosity. It aligns with previous theoretical studies in DPD systems~\cite{Marsh1997a,Groot1997,Chaudhri2010}, where the shear viscosity is decomposed into kinetic and dynamic contributions, that, in the high friction limit, the dynamic component dominates over the kinetic part. This is also consistent with the stress decomposition introduced in Section~\ref{subsubsec:Transport Shear viscosity}, where the dissipative stress tensor gradually outweighs the conservative stress tensor as $\gamma$ increases. It is worth mentioning that the simulations in~\cite{Chaudhri2010} were performed in an idealised setting without the conservative force. As a result, their computed shear viscosity decreases in the small friction limit, which differs from our observations due to the inclusion of the conservative force in our systems. In addition, by switching off the conservative force, we again have verified our results in the case of ideal fluids---they are in good agreement with those obtained in~\cite{Chaudhri2010}.

\begin{figure}[tbp]
  \centering
  \includegraphics[scale=0.38]{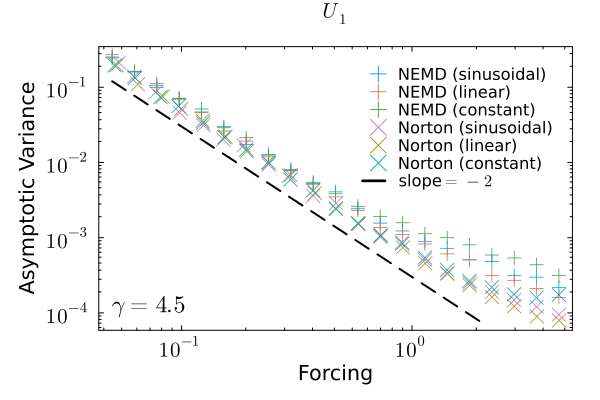}
  \includegraphics[scale=0.38]{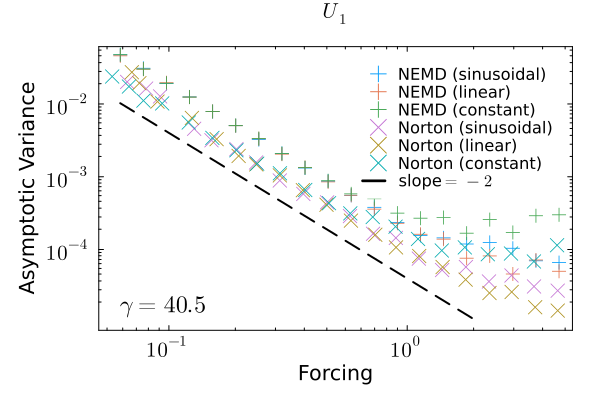}
  \caption{\small
  Double logarithmic plot of the asymptotic variance for the Fourier response $U_1$ against the forcing magnitude for the DPD-NEMD~\eqref{eq:DPD NEMD} and DPD-Norton~\eqref{eq:DPD Norton} dynamics, with three types of external forces (the sinusoidal, piecewise linear, and piecewise constant forces), with friction coefficients of 
  %$\gamma=1$ (top left), 
  $\gamma=4.5$ (left) and $\gamma=40.5$ (right). The expected scaling line for small values of the forcing or response is plotted in dashed lines in all cases.}
  \label{fig:shear viscosity var}
\end{figure}

Furthermore, we compare the asymptotic variance of the Fourier response $U_1$ against the forcing magnitude by using the block averaging method~\cite{Flyvbjerg1989}, with three types of external forces and friction coefficients of $\gamma=4.5$ and $\gamma=40.5$, as shown in Figure~\ref{fig:shear viscosity var}. It can be clearly seen that the asymptotic variances of $U_1$ scale as $\eta^{-2}$ for the DPD-NEMD dynamics and $r^{-2}$ for the DPD-Norton dynamics in the linear regime. With 
%a friction coefficient of $\gamma = 1$, the asymptotic variances of both dynamics are very similar to each other when the forcing magnitude is not over 10, while with 
a friction coefficient of $\gamma = 4.5$, the asymptotic variance associated with the DPD-Norton dynamics appears to be constantly better than that of the DPD-NEMD dynamics, particularly when the forcing magnitude is relatively large. Moreover, the DPD-Norton dynamics clearly outperforms the DPD-NEMD dynamics with a friction coefficient of $\gamma = 40.5$.
%As $\gamma$ increases, the asymptotic variances obtained from the DPD-Norton dynamics appear to be smaller than those from the DPD-NEMD dynamics. 
Unlike the mobility case, the range over which the variance retains the expected order of $-2$ decay remains nearly unchanged with different values of $\gamma$. Regarding the effect of the types of forces, the performance among the three types of forces are almost indistinguishable for both the DPD-NEMD and DPD-Norton dynamics, particularly in the linear regime. 
%In contrast, for the DPD-Norton dynamics, the \corr{sinusoidal}{piecewise linear} force consistently yields the lowest variance, while the piecewise constant force shows the highest variance and begins to deviate from the order $-2$ scaling with smaller forcing magnitudes compared to the other two types of forces.

\section{Conclusions}
\label{sec:Conclusions}

We have reviewed the mathematical formulations of the dissipative particle dynamics (DPD), and have introduced its variants with nonequilibrium molecular dynamics (NEMD) and stochastic Norton dynamics. We have discussed the numerical methods for both the DPD-NEMD and DPD-Norton dynamics. Particularly, we have proposed a numerical method for the DPD-Norton dynamics. The proposed ABOBA method is based on splitting the vector field of the system into `A', `B', and `O' three elementary dynamics, each of which individually preserves a response observable. Therefore, overall the ABOBA method preserves the response observable by construction. We have also proposed an alternative approach for the computation of the shear viscosity in DPD systems. By introducing a spatially dependent external force and constructing the velocity and stress observables in DPD, we have derived a Poisson-type equation that connects the average longitudinal velocity to the applied force, where the shear viscosity can be obtained from the Fourier coefficient of the associated averaged quantities. Furthermore, we have demonstrated in our numerical experiments that the DPD-Norton dynamics can indeed act as an alternative approach for the computation of transport coefficients, including the mobility and the shear viscosity, as the DPD-NEMD dynamics, generalising the results in a recent study in Langevin dynamics~\cite{Blassel2023}. 

To be more precise, in the computation of the mobility, the responses profiles for both the DPD-NEMD and DPD-Norton dynamics coincide in both linear and nonlinear (even for extreme values of the forcing magnitude) regimes in the colour drifts case, while the estimated mobility obtained from the DPD-Norton dynamics is closer to the reference value (obtained from the MSD method) than that from the DPD-NEMD dynamics in the two drifts case. Moreover, we have observed that the colour drifts approach is able to maintain a linear response regime for a wider range of the values of the forcing magnitude than the two drifts approach for both the DPD-NEMD and DPD-Norton dynamics. We have also numerically verified that the corresponding asymptotic variance, which is a key metric to measure the associated computational costs, indeed scales as $\eta^{-2}$ for the DPD-NEMD dynamics and $r^{-2}$ for the DPD-Norton dynamics, at least in the small forcing regime when the friction coefficient is relatively small (larger forcing magnitudes may be needed in the higher friction coefficient limits). The asymptotic variances from the DPD-NEMD dynamics are almost indistinguishable between the two drifts and colour drifts approaches, while those from the DPD-Norton dynamics are generally smaller, especially for larger values of $\gamma$. Notably, with $\gamma = 40.5$, the asymptotic variance from DPD-Norton is significantly reduced compared to that from DPD-NEMD, and the regime with the expected order of $-2$ scaling extends to larger forcing magnitudes.
%Particularly, with a friction coefficient of $\gamma=4.5$, the asymptotic variances obtained from the DPD-Norton dynamics are almost one magnitude smaller than those obtained from the DPD-NEMD dynamics when the value of the forcing magnitude is small.

Meanwhile, we have carried out a thorough numerical investigation for the computation of the shear viscosity by using both the DPD-NEMD and DPD-Norton dynamics. The methodology is based on the relation between the spatially dependent external force and the Fourier response of the average longitudinal velocity as presented in~\eqref{eq:shear viscosity computation}. In our numerical experiments, the sinusoidal, piecewise linear, and piecewise constant forces have been considered, where both dynamics generate nearly identical results in a wide range of the values of the forcing magnitude, including both linear and nonlinear regimes. We have further verified that the estimator $F_1 / U_1$ converges as a function of the system size $N$ with a rate asymptotically consistent with the theoretical expression in~\eqref{eq:shear viscosity computation} of order $N^{-2/3}$. In addition, the histograms of the relevant observables ($R$ and $\lambda$) in equilibrium become increasingly concentrated around zero as $N$ increases, with variances scaling as $N^{-1}$ for the DPD-NEMD dynamics and $N^{-5/3}$ for the DPD-Norton dynamics, reflecting stronger concentration in the Norton formulation. Moreover, the computed shear viscosity increases monotonically with the friction coefficient $\gamma$, consistent with previous theoretical studies showing the dominance of the dynamic contribution in the high friction limit. Lastly, we have compared the asymptotic variances of $U_1$ with different types of forces and different values of $\gamma$, and have observed that both the DPD-NEMD and DPD-Norton dynamics are relatively insensitive to the types of forces in the linear regime. Moreover, as $\gamma$ increases, the asymptotic variances obtained from the DPD-Norton dynamics appear to be smaller than those from the DPD-NEMD dynamics.

%It is worth mentioning that~\eqref{eq:shear} admits analytical solutions for any smooth function $F(y)$, which provides an alternative approach for the computation of the shear viscosity. However, for those nonsmooth functions, such as the piecewise linear and piecewise constant forces, the Fourier transform used in this article remains applicable estimation of the shear viscosity.  

Furthermore, it will be of interest to extend the present analysis to further investigate how the DPD-Norton dynamics performs with more complicated potential energies, for instance, in polymer melts~\cite{Shang2017}.
%A natural extension of the current work is to generalise the results on the computation of other transport properties to DPD. However, it is nontrivial and we therefore leave it for future work. It will also be of interest to explore how the DPD-Norton dynamics would perform with more complicated potential energies, for instance, in polymer melts~\cite{Shang2017}.

\section*{CRediT authorship contribution statement}

\textbf{Xinyi Wu:} Conceptualization, Methodology, Software, Validation, Formal analysis, Investigation, Writing - original Draft, Writing - review \& editing, Visualization. \textbf{Xiaocheng Shang:} Conceptualization, Methodology, Validation, Formal analysis, Investigation, Writing - original draft, Writing - review \& editing, Supervision, Funding acquisition.

\section*{Declaration of competing interest}

The authors declare that they have no known competing financial interests or personal relationships that could have appeared to influence the work reported in this paper.

\section*{Acknowledgements}

The authors thank No\'{e} Blassel, Gabriel Stoltz, Urbain Vaes, and the anonymous referees for their valuable suggestions and comments. XW acknowledges the support from the University of Birmingham and the China Scholarship Council. XS acknowledges the support of the Royal Society through the International Exchanges Scheme (reference number IES$\backslash$R3$\backslash$203007), the Royal Society through the Research Grants (reference number RG$\backslash$R2$\backslash$232257), the Isaac Newton Institute for Mathematical Sciences through the Network Support for the Mathematical Sciences initiative (EPSRC reference number EP/V521929/1), the London Mathematical Society through the Emmy Noether Fellowship (reference number EN-2223-09), and CECAM and CCP5 via their CECAM/CCP5 sandpits. This work was supported by Institute of Mathematics for Industry, Joint Usage/Research Center in Kyushu University. (FY2025 Workshop(I) ``Workshop on Mathematics for Machine Learning and Its Application to Industry'' (2025a027).)

\appendix

%\begin{appendices}
\section{Proof of Proposition~\ref{prop1}}
\label{appendix:prop1}
Re-decompose the generator of a perturbed DPD system for the equilibrium and nonequilibrium parts as follows:
\begin{equation}
  \Lg = \Lg_{0} + \Lg_{\eta} \, ,
\end{equation}
where the generator of the equilibrium part $\Lg_{0}$ consists of Hamiltonian and thermostat components, 
\begin{equation}
  \Lg_{0}=\Lg_{\mathrm{ham}}+\Lg_{\mathrm{thm}}\, ,
\end{equation}
where
\begin{subequations}
\begin{align}
    \Lg_{\mathrm{ham}} &= \M^{-1}\p\cdot\nabla_\q-\nabla U(\q)\cdot\nabla_\p\, ,\\
    \Lg_{\mathrm{thm}} &=-\gamma\GammaB(\q)\M^{-1}\p\cdot\nabla_\p+\frac{\sigma^2}{2}\SigmaB(\q)[\SigmaB(\q))]^T:\nabla^2_\p\, ,
\end{align}
\end{subequations}
and the generator of the nonequilibrium perturbations $\Lg_{\eta}$ is given by
\begin{equation}
  \Lg_{\eta} = \eta\sum\limits_{i=1}^{N}F\left(q^{y}_{i}\right)\partial_{p^{x}_{i}} \, .
\end{equation}
%are defined as follows ($\Lg_0$ and $\Lg_{\eta}$ are similar to the notations of $\mathcal{A}_{0}$ and $\mathcal{B}$ in~\cite{Joubaud2012}, but in the case of DPD):
The corresponding adjoint operator is given by
\begin{equation}
    \Lg^*=\Lg^*_{0}+\Lg_{\eta}^* \, ,
\end{equation}
where
\begin{subequations}
\begin{align}
  \Lg^*_{0} &= -\Lg_{\mathrm{ham}} + \Lg_{\mathrm{thm}}\, ,\\
  \Lg_{\eta}^* &= -\eta\sum\limits_{i=1}^{N}\left(F\left(q^{y}_{i}\right)\partial_{p^{x}_{i}}-\frac{\beta}{m}p^{x}_{i}F\left(q^{y}_{i}\right)\right)\, .
\end{align}
\end{subequations}
The proof of equations~\eqref{eq:hamF} below in the context of DPD is similar to the proof of Corollary 1 in~\cite{Joubaud2012} in Langevin dynamics:
\begin{equation}
  \lim\limits_{\eta \rightarrow 0} \frac{\left\langle\Lg_{0}U_x^\varepsilon(Y,\q,\p)\right\rangle_{\eta}}{\eta} = -\frac{\beta}{m}\left\langle U_x^\varepsilon(Y,\q,\p), \sum\limits_{i=1}^{N}p^{x}_{i}F\left(q^{y}_{i}\right) \right\rangle_{L^2(\rho_{\beta})} \, ,\label{eq:hamF}
\end{equation} 
where $\langle f,g \rangle_{L^2(\rho_{\beta})}$ denotes the scalar product of the observable functions $f$ and $g$, associated with the measure of the dynamics~\eqref{eq:DPD}. Therefore, we have
\begin{equation}
  \lim\limits_{\varepsilon \rightarrow 0} \lim\limits_{\eta \rightarrow 0} \frac{\left\langle\Lg_{0}U_x^\varepsilon(Y,\q,\p)\right\rangle_{\eta}}{\eta} = -\frac{1}{m}F(Y) \, .
\end{equation}
Meanwhile, splitting $\Lg_{0}U_x^\varepsilon(Y,\q,\p)$ into Hamiltonian and thermostat parts, we have
\begin{subequations}
\begin{align}
  \Lg_{\mathrm{ham}}U_x^\varepsilon(Y,\q,\p) &= -\frac{L_y}{Nm}\frac{\dd}{\dd Y}\left(\sum\limits_{i=1}^N\frac{p^{x}_{i}p^{y}_{i}}{m}\chi_\varepsilon\left(q^{y}_{i}-Y\right)-\sum\limits_{1\leq i<j\leq N}U'\left(r_{ij}\right)e_{ij}^x\int_{q^{y}_{j}}^{q^{y}_{i}}\chi_\varepsilon(s-Y) \, \dd s\right)\, ,\label{eq:ham}\\
  \Lg_{\mathrm{thm}}U_x^\varepsilon(Y,\q,\p) &= -\frac{\gamma L_y}{Nm}\sum\limits_{1\leq i<j\leq N}\omega^{\D}\left(r_{ij}\right)\left(\mathbf{e}_{ij}\cdot\mathbf{v}_{ij}\right)e_{ij}^x\left[\chi_\varepsilon\left(q^{y}_{i}-Y\right)-\chi_\varepsilon(q^{y}_{j}-Y)\right] \nonumber \\ 
  &= \frac{\gamma L_y}{Nm}\frac{\dd}{\dd Y}\left(\sum\limits_{1\leq i<j\leq N}\omega^{\D}\left(r_{ij}\right)\left(\mathbf{e}_{ij}\cdot\mathbf{v}_{ij}\right)e_{ij}^x\int_{q^{y}_{j}}^{q^{y}_{i}}\chi_\varepsilon(s-Y) \, \dd s\right) \, .
\end{align}
\end{subequations} 
Recalling the definition of the $xy$ component of stress tensor in~\eqref{eq:stress}, for the partial derivative of $\Sigma_{xy}$ with respect to $Y$, we have
\begin{equation}
  -\rho m\Lg_{0}U_x^\varepsilon(Y,\q,\p) = \frac{\partial\Sigma_{xy}^\varepsilon(Y,\q,\p)}{\partial Y} \, .
\end{equation}
Therefore, passing to the limits $\varepsilon \rightarrow 0$ and $\eta \rightarrow 0$, we have 
\begin{equation}
    \rho F(Y)=\frac{\partial\sigma_{xy}(Y)}{\partial Y}\, .
\end{equation}

\bibliographystyle{elsarticle-num}
% \bibliographystyle{elsarticle-harv}
% \bibliographystyle{elsarticle-num-names}
% \bibliographystyle{model1a-num-names}
% \bibliographystyle{model1b-num-names}
% \bibliographystyle{model1c-num-names}
% \bibliographystyle{model2-names}
% \bibliographystyle{model3a-num-names}
% \bibliographystyle{model3-num-names}
% \bibliographystyle{model4-names}
% \bibliographystyle{model5-names}
% \bibliographystyle{model6-num-names}

%\bibliographystyle{plain}
%\bibliographystyle{is-abbrv}
%\bibliographystyle{unsrt}

%\bibliography{sample}
\bibliography{refs}

\end{document}